\documentclass[lettersize,journal]{IEEEtran}
\usepackage{amsmath,amsfonts}
\usepackage{algorithmic}
\usepackage{algorithm}
\usepackage{array}
\usepackage{textcomp}
\usepackage{stfloats}
\usepackage{url}
\usepackage{verbatim}
\usepackage{graphicx}

\usepackage{subcaption}
\usepackage{cite}
\usepackage{amsmath,amssymb,amsfonts}
\usepackage{algorithmic}
\usepackage{graphicx}
\usepackage{textcomp}
\usepackage{xcolor}
\usepackage{multirow}
\usepackage{csquotes}

\usepackage{tikz}
\usepackage{circuitikz}
\usepackage{pgfplots}
\usetikzlibrary{patterns}
\usetikzlibrary{backgrounds}
\usetikzlibrary{calc}
\usepackage{tabularx, booktabs} 

\usepackage{float}
\usepackage{bbm}
\usepackage{listings}
\usepackage{asn1}
\lstdefinestyle{3gppmsg}{
    language=asn1,
    basicstyle=\footnotesize,
    showstringspaces=false,
    breaklines=true,
    frame=top,
    frame=bottom,
    aboveskip=1.5em,
    belowskip=1.5em,
    captionpos=t
}

\def \fwidth{0.7\linewidth}
\def \fheight{0.45\linewidth}

\def \singfwidth{0.55\linewidth}
\def \singfheight{0.35\linewidth}

\colorlet{mycolor1}{cyan}
\colorlet{mycolor2}{orange}
\colorlet{mycolor3}{violet}
\colorlet{mycolor4}{red}
\colorlet{mycolor5}{blue}

\newcolumntype{M}[1]{>{\centering\arraybackslash}m{#1}}

\usepackage[acronyms]{glossaries}
\newacronym{cdf}{CDF}{Cumulative Distribution Function}
\newacronym{nlos}{NLoS}{Non Line of Sight}
\newacronym{los}{LoS}{Line of Sight}
\newacronym{hpbw}{HPBW}{Half Power Beam Width}
\newacronym{tx1}{TX1}{Transmitter 1}
\newacronym{tx2}{TX2}{Transmitter 2}
\newacronym[plural=SNRs,firstplural=Signal to Noise Ratios]{snr}{SNR}{Singal to Noise Ratio}
\newacronym{ue}{UE}{User Equipment}
\newacronym{6g}{6G}{Sixth Generation}
\newacronym{vr}{VR}{Virtual Reality}
\newacronym{harq}{HARQ}{Hybrid Automatic Retransmission reQuest}
\newacronym{csi}{CSI}{Channel State Information}
\newacronym{capex}{CAPEX}{Capital Expenditure}
\newacronym{opex}{OPEX}{Operational Expenditure}
\newacronym{aoa}{AoA}{Angle of Arrival}
\newacronym{toa}{ToA}{Time of Arrival}
\newacronym[plural=ADCs,firstplural=Analog to Digital Converters]{adc}{ADC}{Analog to Digital Converter}
\newacronym[plural=TRNs,firstplural=Training Fields]{trn}{TRN}{Training Field}
\newacronym{lpf}{LPF}{Low Pass Filter}
\newacronym{music}{MUSIC}{MUltiple SIgnal Classification}
\newacronym{2dmusic}{2D-MUSIC}{2D MUltiple SIgnal Classification}
\newacronym{mimo}{MIMO}{Multiple Input Multiple Output}
\newacronym{mumimo}{MU-MIMO}{Multi User Multiple Input Multiple Output}
\newacronym{simo}{SIMO}{Single Input Multiple Output}
\newacronym{cir}{CIR}{Channel Impulse Response}
\newacronym{cfr}{CFR}{Channel Frequency Response}
\newacronym{mrc}{MRC}{Maximum Ratio Combining}
\newacronym[plural=BSs,firstplural=Base Stations]{bs}{BS}{Base Station}
\newacronym{tdd}{TDD}{Time Division Duplexing}
\newacronym{ofdm}{OFDM}{Orthogonal Frequency Division Multiplexing}
\newacronym{af}{AF}{Array Factor}
\newacronym{tx}{TX}{Transmitter}
\newacronym{rx}{RX}{Receiver}
\newacronym{rf}{RF}{Radio Frequency}
\newacronym{mu}{MU}{Multi User}
\newacronym{rms}{RMS}{Root Mean Square}
\newacronym{fft}{FFT}{Fast Fourier Transform}
\newacronym{wlan}{WLAN}{wireless local area network}
\newacronym{ofdma}{OFDMA}{Orthogonal Frequency Division Multiplexing Access}
\newacronym{jcs}{JCAS}{Joint Communication and Sensing}
\newacronym{pmcw}{PMCW}{phase modulated continuous wave}
\newacronym{mmwave}{mmWave}{millimeter wave}
\newacronym{rmse}{RMSE}{Root Mean Square Error}
\newacronym{iir}{IIR}{Infinite Impulse Response}
\newacronym{LFM}{LFM}{linear frequency modulated}
\newacronym{FMCW}{FMCW}{frequency modulated continuous wave}
\newacronym{PPM}{PPM}{pulse position modulated}
\newacronym{eurllc}{eURLLC}{extreme Ultra Reliable Low Latency Communications}
\newacronym{pec}{PEC}{Perfect Electric Conductor}
\newacronym{cst}{CST}{CST Microwave Studio}
\newacronym{svd}{SVD}{Singular Value Decomposition}
\newacronym[plural=KPIs,firstplural=Key Performance Indicators]{kpi}{KPI}{Key Performance Indicator}
\newacronym{sir}{SIR}{Signal to Interference Ratio}
\newacronym{pdf}{PDF}{Probability Density Function}
\newacronym[longplural={Uniform Linear antenna Array}, shortplural={ULAs}]{ula}{ULA}{Uniform Linear antenna Array}
\newacronym{fov}{FoV}{Field of View}
\newacronym[longplural={Random Variables}, shortplural={r.v.s}]{rv}{r.v.}{Random Variable}
\newacronym{urllc}{URLLC}{Ultra Reliable Low Latency Communications}
\newacronym{fspl}{FSPL}{Free Space Path Loss}
\newacronym{dtft}{DTFT}{Discrete Time Fourier Transform}
\newacronym{vga}{VGA}{Variable Gain Amplifier}
\newacronym{lo}{LO}{Local Oscillator}
\newacronym[plural=ICs,firstplural=Integrated Circuits]{ic}{IC}{Integrated Circuit}
\newacronym[longplural={Resource Blocks}, shortplural={RBs}]{rb}{RB}{Resource Block}
\newacronym{mac}{MAC}{Medium Access Control}
\newacronym{se}{SE}{Spectral Efficiency}
\newacronym[plural=LNAs,firstplural=Low Noise Amplifiers]{lna}{LNA}{Low Noise Amplifier}
\newacronym{pf}{PF}{Proportional Fairness}
\newacronym{dci}{DCI}{Downlink Control Information}
\newacronym{rrc}{RRC}{Radio Resource Control}
\newacronym{3gpp}{3GPP}{3rd Generation Partnership Project}
\newacronym{asn1}{ASN.1}{Abstract Syntax Notation One}
\newacronym{ie}{IE}{Information Element}
\newacronym{bwp}{BWP}{BandWidth Part}
\newacronym{mcs}{MCS}{Modulation and Coding Scheme}
\newacronym{mmimo}{mMIMO}{massive Multiple Input Multiple Output}
\newacronym{ssb}{SSB}{Synchronization Signal Block}
\newacronym{pdcch}{PDCCH}{Physical Downlink Control Channel}
\newacronym{ul}{UL}{Uplink}
\newacronym{dl}{DL}{Downlink}

\newcommand{\edit}[1]{\textcolor{black}{#1}}

\newcommand{\msgname}{Msg.~}

\newcommand{\E}{\mathrm{E}}

\usepackage{enumitem}
\newlist{taskenum}{enumerate}{1}
\setlist[taskenum]{label=\textit{T.\arabic*}}

\def\BibTeX{{\rm B\kern-.05em{\sc i\kern-.025em b}\kern-.08em
    T\kern-.1667em\lower.7ex\hbox{E}\kern-.125emX}}

\usetikzlibrary {patterns.meta}
\tikzdeclarepattern{
  name=mylines,
  parameters={
      \pgfkeysvalueof{/pgf/pattern keys/size},
      \pgfkeysvalueof{/pgf/pattern keys/angle},
      \pgfkeysvalueof{/pgf/pattern keys/line width},
  },
  bounding box={
    (0,-0.5*\pgfkeysvalueof{/pgf/pattern keys/line width}) and
    (\pgfkeysvalueof{/pgf/pattern keys/size},
0.5*\pgfkeysvalueof{/pgf/pattern keys/line width})},
  tile size={(\pgfkeysvalueof{/pgf/pattern keys/size},
\pgfkeysvalueof{/pgf/pattern keys/size})},
  tile transformation={rotate=\pgfkeysvalueof{/pgf/pattern keys/angle}},
  defaults={
    size/.initial=5pt,
    angle/.initial=45,
    line width/.initial=.4pt,
  },
  code={
      \draw [line width=\pgfkeysvalueof{/pgf/pattern keys/line width}]
        (0,0) -- (\pgfkeysvalueof{/pgf/pattern keys/size},0);
  },
}

\begin{document}

\title{Heterogeneous Rank Beamforming for Industrial Communications}

\author{Andrea Bedin,~\IEEEmembership{Graduate Student Member,~IEEE}, Akshay Jain,~\IEEEmembership{Member,~IEEE}, Andrea Zanella,~\IEEEmembership{Senior Member,~IEEE}, Karthik Upadhya,~\IEEEmembership{Member,~IEEE}
\thanks{A. Bedin  (corresponding author, andrea.bedin.2@phd.unipd.it), A. Jain (akshay.2.jain@nokia-bell-labs.com) and K. Upadhya (karthik.upadhya@nokia-bell-labs.com) are with Nokia Bell Labs, Espoo, Finland. A. Bedin and A. Zanella (andrea.zanella@unipd.it) are with the Department of Information Engineering, University of Padova, Italy.  This work has received funding from the European Union’s EU Framework Programme for Research and Innovation Horizon 2020 under Grant Agreement No 861222.}}

\maketitle

\begin{abstract}
This paper proposes a novel hardware beamforming architecture, which is capable of utilizing a different number of \gls{rf} chains in different parts of the bandwidth. It also shows that a proportional fairness scheduler will effectively utilize the high rank part of the bandwidth in a multi-user setting, thus operating more efficiently and effectively than classical beamforming schemes.
\end{abstract}


\glsresetall

\section{Introduction}
Wireless communication systems have emerged as a fundamental enabler for Industry 4.0, promising rapid transformation of traditional industrial environments into smart and interconnected ecosystems. Specifically, the integration of wireless technologies holds the potential to revolutionize the way industrial processes are managed, monitored, and optimized. Enabling real-time data exchange between pieces of industrial equipment, as well as remote control and maintenance of the machines by the operators, the use of wireless communication in industrial environments provides several advantages, such as reduced costs, improved productivity, and increased flexibility. Concretely, applications such as Digital twins, Collaborative robots and telepresence are expected to be serviced through wireless communications. 

Tab.~\ref{tab:requirements} lists the requirements for Digital twins and cooperative robots applications. This list reports the values defined by the European \gls{6g} flagship project Hexa-X \cite{hexa_71}. The latency requirements for these applications can reach as low as sub-millisecond values, which is highly challenging to achieve in traditional wireless networks. This tight requirement is motivated by the fact that digital twins and cooperative robots require real-time interaction and decision-making capabilities. Any communication delay can lead to safety hazards and performance degradation. In addition, the reliability requirements for these applications are also very stringent, with an error rate as low as $10^{-9}$. This level of reliability is necessary to ensure that the robots operate accurately and safely without causing any harm to the environment or humans.

While these requirements might seem extreme for today's factories, where robots typically move at limited speeds, they are crucial for enabling much faster operations while maintaining safety and control in future manufacturing plans. Tab.~\ref{tab:requirements_telepresence} reports a more detailed list of the \gls{ul} and \gls{dl} requirements for telepresence and \gls{vr}, as for the Hexa-X project \cite{hexa_71}. Here we can observe the heterogeneity of the requirements for different traffic categories within the same service. For example, we can observe that the haptic interactions, which require delays as low as $1$ms but are not highly demanding in terms of bitrate, must coexist with video streaming, which is, instead, very bandwidth-intensive. While this coexistence could be handled by deploying multiple networks with different capabilities, this would entail much higher \gls{capex} and \gls{opex} for the network operator. Therefore, it is crucial from an economic standpoint to meet all of these requirements in a single network.

\begin{table}[t]
\centering
\caption{Requirements for selected applications}
\begin{tabular}{c|c|c|c}
\hline
 \textbf{Use Case} & \textbf{Latency} & \textbf{Reliability} & \textbf{Data Rate} \\ \hline
\edit{Cooperative} Robots & $1$-$50$~ms & $10^{-9}$ & kbps \\
Digital Twins & $0.1$ - $100$~ms & $10^{-2}$ - $10^{-6}$ & $1$-$10$~Gbps 
\end{tabular}
\label{tab:requirements}
\end{table}

\begin{table}[t]
\centering
\caption{Telepresence requirements}
\begin{tabular}{c|c|c|c|c}
\hline
\textbf{Use Case} & \textbf{Traffic type} & \textbf{Latency} & \textbf{Reliability} & \textbf{Data Rate} \\ \hline
Teleoperation & Haptic UL & $5$~ms & $10^{-1}$ & $1$-$4$~kbps \\ 
			  & Haptic DL & $1$-$50$~ms & $10^{-1}$ & $1$-$4$~kbps \\ 
 			  & Audio UL & $10$~ms & $10^{-1}$ & $5$-$512$~kbps \\ 
 			  & Video UL & $10$~ms & $10^{-3}$ & $1$-$100$~Mbps \\ \hline
Immersive VR  & Haptic UL & $1$-$10$~ms & $10^{-3}$ & $1$-$4$~kbps \\ 
			  & Haptic DL & $1$-$10$~ms & $10^{-3}$ & $1$-$4$~kbps \\ 
 			  & Audio UL & $20$~ms & $10^{-1}$ & $5$-$512$~kbps \\ 
 			  & Video UL & $10$-$20$~ms & $10^{-3}$ & $1$-$100$~Mbps \\

\end{tabular}
\label{tab:requirements_telepresence}
\end{table}

In an effort to meet increasing demand for higher data rates, lower latency and reliable communication in such challenging environments, wireless communication technology has evolved significantly over the years. Many studies have been carried out on different aspects such as \gls{mmimo}, beamforming techniques, \gls{mac}, etc \cite{industry_coding, industry_standards}. Of particular interest for this work are the beamforming techniques \cite{beamforming_survey, beamforming_survey_2, beamforming_survey_3, beamforming_hybrid_vs_digital}. In recent years, there has been a focus on hybrid and analog beamforming strategies, which are oriented towards sparse channels such as those observed in indoor environments \cite{cluster_number}. However, the channel observed in industrial environments is different, given the very complex and rich multipath \cite{factory_28G, factory_28G_2}. Therefore the reliability of existing wireless systems in such environments is degraded, and their ability to exploit the rich multipath propagation is severely limited. On the other hand, due to the large number of antennas needed to overcome the path loss of \gls{mmwave}, the choice of moving towards analog or hybrid beamforming, motivated by the complexity and power consumption of \edit{the \glspl{adc} required to build digital beamforming devices}, might seem inevitable. We therefore observe a clear need for an alternative solution that is capable of effectively exploiting the highly rich multipath channel, as well as maintaining manageable costs and complexity.

Addressing this ambitious objective, in this paper we propose:
\begin{itemize}
\item A novel hardware architecture \edit{\cite{arch_patent}}, described in Sec.~\ref{sec:architectures}, that supports the trade-off between bandwidth and rank of the channel. Such an architecture, as discussed in Sec.~\ref{sec:arch_compl_cost}, is slightly more complex than the classical hybrid beamforming architecture, but still significantly simpler, cheaper and energy efficient compared to fully digital beamforming architecture. This aims to overcome the limitations of analog beamforming in at least part of the bandwidth with limited increase in costs.
\item An analysis of the performance of \gls{pf} resource allocation for multiple users equipped with the proposed architecture, presented in Sec.~\ref{sec:mimo_sched}, showing that in a multi-user environment a large part of the communication can be performed exploiting digital beamforming, even if the users do not operate with digital beamforming on the \textit{full} band. 
\item An update to the \gls{3gpp} signaling, described in Sec.~\ref{sec:3gpp}, that enable the support of this technology, with a small amount of additional parameters in the \gls{ue} configuration and resource allocation messages.
\end{itemize} 

\section{Beamforming architectures} \label{sec:architectures}

\subsection{Classical beamforming architectures} \label{sec:limitations}
\begin{figure}[t]
\centering
\resizebox{0.7\linewidth}{!}{\begin{tikzpicture}

\def \apos {-7}
\def \ashift {0.5}
\def \posend {-13}
\def \shiftapos {\apos + 0.5}
\def \mixpos {-10}
\def \shiftmixpos {\mixpos - 0.5}
\def \dmixpos {-4.5}
\def \ampendpos {-3}
\def \antpos {0}
\def \donepos {-2.5}
\def \dtwopos {-3}
\def \dthreepos {-3.5}
\def \lineone {0}
\def \linetwo {-1.2}
\def \linethree {-3.6}

\def \aone {\lineone + \ashift}
\def \atwo {\linetwo + \ashift}

\def \athree  {\linethree + \ashift}

\def \dlineone {-5.1}
\def \dlinetwo {-6.37}
\def \dlinethree {-8.9}

\def \posouttext {\posend - 0.8}

\def \chaintoshift {0.15}

\draw


node[adder, fill=yellow!30!white] at (\apos -1,\linetwo - \chaintoshift) (add2) {}

(\donepos,\lineone) to[short, *-] node[short]{} (\donepos,\lineone - \chaintoshift)
to[short] node[short]{} (\donepos-1.95,\lineone - \chaintoshift)
to [phaseshifter, fill=yellow!30!white] node[short]{}(\apos+0.75,\lineone - \chaintoshift)
to[short] node[short]{} (\apos+0.75,\lineone - \chaintoshift - 0.5)
to[short] node[short]{} (\apos-1,\lineone - \chaintoshift - 0.5)
to[short] node[short]{} (add2.north)

(\dtwopos,\linetwo) to[short, *-] node[short]{} (\dtwopos,\linetwo - \chaintoshift)
to[short] node[short]{} (\dtwopos-0.25,\linetwo - \chaintoshift)
to [phaseshifter, fill=yellow!30!white] node[short]{}(add2.east)

(\dthreepos,\linethree) to[short, *-] node[short]{} (\dthreepos,\linethree - \chaintoshift)
to[short] node[short]{} (\dthreepos,\linethree - 0.2)
to [phaseshifter, fill=yellow!30!white] node[short]{} (\apos -0.3,\linethree - \chaintoshift)
to[short] node[short]{} (\apos -1,\linethree - \chaintoshift)
to[short] node[short]{} (add2.south)


node[adder, fill=yellow!30!white] at (\apos,0) (add) {}

node[bareantenna] (ant1) at (\antpos , \aone) {}
(ant1.center)  to node[short]{} (\antpos,  \lineone)
to [amp, fill=blue!30!white] node[short]{}(\ampendpos,\lineone) 
to node[short]{}(-4,\lineone) 
to [phaseshifter, fill=yellow!30!white] (add.east) 

node[bareantenna] (ant2)  at (\antpos + 0.5, \atwo) {}
(ant2.center)  to node[short]{} (\antpos+ 0.5,  \linetwo)
to [amp, fill=blue!30!white] node[short]{}(\ampendpos,\linetwo) 
to node[short]{}(-4,\linetwo) 
to [phaseshifter, fill=yellow!30!white] node[short]{}(\shiftapos,\linetwo)
to node[short]{}( \apos,\linetwo)
 to (add.south)
 
 node[bareantenna] (ant3)  at (\antpos + 1.5,  \athree) {}
(ant3.center)  to node[short]{} (\antpos+ 1.5,  \linethree)
to [amp, fill=blue!30!white] node[short]{}(\ampendpos, \linethree) 
to  node[short]{}(-4, \linethree) 
to [phaseshifter, fill=yellow!30!white] node[short]{}(\shiftapos, \linethree)
to node[short]{}( \apos, \linethree)
 to (add.south)
 
 node[mixer, fill=yellow!30!white] at (\mixpos,0) (mixwb) {}
 (add.west) to (mixwb.east)
 (mixwb.west) to[adc, -o, fill=yellow!30!white] node[short](analogend){}(\posend,0)

  node[mixer, fill=yellow!30!white] at (\mixpos,\linetwo - \chaintoshift) (mixwb2) {}
 (add2.west) to (mixwb2.east)
 (mixwb2.west) to[adc, -o, fill=yellow!30!white] node[short](analogend){}(\posend,\linetwo - \chaintoshift)
 

 ;

 \filldraw[color=black, fill=black, thick] (0.25,-1.8) circle (1pt) ;
\filldraw[color=black, fill=black, thick] (0.5,-2.4) circle (1pt) ;
 \filldraw[color=black, fill=black, thick] (0.75,-3) circle (1pt) ;

\filldraw[color=black, fill=black, very thick] (-4,-2) circle (1pt) ;
\filldraw[color=black, fill=black, very thick] (-4,-2.5) circle (1pt) ;
 \filldraw[color=black, fill=black, very thick] (-4,-3) circle (1pt) ;
 
 

\begin{scope}[on background layer]
	

	\draw[yellow!80!black, dashed, line width = 2] (-12.75, -4.3) -- (-2.25, -4.3) -- (-2.25,1) -- (-12.75, 1) -- cycle;
	
	\fill[yellow!10!white] (-12.75, -4.3) -- (-2.25, -4.3) -- (-2.25, 1) -- (-12.75, 1);
\end{scope}


 
\end{tikzpicture}}
\caption{Classical fully connected hybrid beamforming hardware architecture. }
\label{fig:architecture_class}
\end{figure}
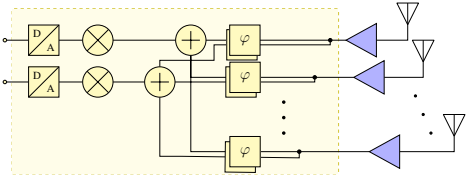

The classical fully connected hybrid beamforming architecture, depicted in in Fig.~\ref{fig:architecture_class} is widely used in the industry for mmWave communications. 

However, this architecture has some limitations that can impact its effectiveness in modern wireless communication systems. One of the main limitations is the need for long and complex beam training and refinement procedures to establish and maintain a connection. Before transmitting the signal, the system needs to determine the optimal beam direction and shape, which involves sweeping through different beam directions and measuring the channel response. This process can take a long time, and it needs to be performed for each user within the system, so that it can become a significant overhead especially when the number of users is large. Additionally, these beam training and refinement procedures can cause excess delay in time-sensitive packets, as communication is not possible while the system is training the beam. Secondly, due to the limitations in \gls{csi} acquisition capability \edit{(i.e., the inability of the system to acquire the full channel matrix, but rather only to measure the power for a specific beam)}, modern wireless communication systems often make use of \edit{a small codebook of} pencil beams \edit{to limit the number of channel measurements \cite{beam_training}. These beams} concentrate the array gain in a single spatial direction \edit{ and are effective} in maximizing throughput for a line of sight scenario with limited reflections, where the signal can travel directly from the transmitter to the receiver \edit{without obstructions and with little multipath}. However, \edit{when operating in an environment with rich multipath,} using pencil beams comes at a cost. By focusing on a single multipath component, pencil beams sacrifice spatial diversity. Hence, when the selected component is blocked by an obstacle, this strategy is likely to experience a complete signal loss, and therefore not be able to communicate until a time consuming beam training is performed. 

\begin{figure}[t]
\centering
\resizebox{0.7\linewidth}{!}{\begin{tikzpicture}

\def \apos {-7}
\def \ashift {0.5}
\def \posend {-13}
\def \shiftapos {\apos + 0.5}
\def \mixpos {-5}
\def \shiftmixpos {\mixpos - 0.5}
\def \ampendpos {-3}
\def \antpos {0}
\def \lineone {0}
\def \linetwo {-1.2}
\def \linethree {-3.6}

\def \aone {\lineone + \ashift}
\def \atwo {\linetwo + \ashift}

\def \athree  {\linethree + \ashift}

\def \posouttext {\posend - 0.8}

\def \chaintoshift {0.15}

\tikzset{m1/.style={muxdemux, muxdemux def={Lh=7, Rh=9, NL=1, NB=0, NR=4}}}

\draw

node[mixer, fill=cyan!30!white] at (\mixpos,\lineone) (mixer1) {}
node[mixer, fill=cyan!30!white] at (\mixpos,\linetwo) (mixer2) {}
node[mixer, fill=cyan!30!white] at (\mixpos,\linethree) (mixer3) {}


node[bareantenna] (ant1) at (\antpos , \aone) {}
(ant1.center)  to node[short]{} (\antpos,  \lineone)
to [amp, fill=blue!30!white] node[short]{}(\ampendpos,\lineone) 
to (mixer1.east) to[adc,fill=cyan!30!white, -o] node[short]{} (\posend, \lineone)

node[bareantenna] (ant2)  at (\antpos + 0.5, \atwo) {}
(ant2.center)  to node[short]{} (\antpos+ 0.5,  \linetwo)
to [amp, fill=blue!30!white] node[short]{}(\ampendpos,\linetwo) 
to (mixer2.east) to[adc,fill=cyan!30!white, -o] node[short]{} (\posend, \linetwo)

 node[bareantenna] (ant3)  at (\antpos + 1.5,  \athree) {}
(ant3.center)  to node[short]{} (\antpos+ 1.5,  \linethree)
to [amp, fill=blue!30!white] node[short]{}(\ampendpos, \linethree) 
to (mixer3.east) to[adc,fill=cyan!30!white, -o] node[short]{} (\posend, \linethree)



 ;

 \filldraw[color=black, fill=black, thick] (0.25,-1.8) circle (1pt) ;
\filldraw[color=black, fill=black, thick] (0.5,-2.4) circle (1pt) ;
 \filldraw[color=black, fill=black, thick] (0.75,-3) circle (1pt) ;

\filldraw[color=black, fill=black, very thick] (-4,-2) circle (1pt) ;
\filldraw[color=black, fill=black, very thick] (-4,-2.5) circle (1pt) ;
 \filldraw[color=black, fill=black, very thick] (-4,-3) circle (1pt) ;

\begin{scope}[on background layer]

	\draw[cyan!80!black, dashed, line width = 2] (-12.75, -4.3) -- (-2.25, -4.3) -- (-2.25,1) -- (-12.75, 1) -- cycle;
	
	\fill[cyan!10!white] (-12.75, -4.3) -- (-2.25, -4.3) -- (-2.25, 1) -- (-12.75, 1);
\end{scope}

\end{tikzpicture}}
\caption{Classical fully digital beamforming hardware architecture. }
\label{fig:architecture_digital}
\end{figure}
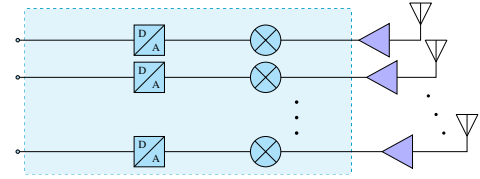

The na\"ive solution to this issue is to use a fully digital beamforming architecture. \edit{The \gls{rf} frontend of such an architecture is depicted in Fig.~\ref{fig:architecture_digital}. Note that, for the sake of a compactness, in the following we will refer to a \gls{rf} frontend that enable the implementation of digital beamforming as a digital beamforming architecture.} This architecture allows the receiver to collect the full signal information for each antenna, and then perform the combining in the digital domain. 
\edit{While the exact digital beamforming algorithms are outside the scope of this paper, it is clear that this scheme} provides some advantages, including but not limited to
\begin{itemize}
\item The ability to combine signals from different antennae with different coefficients at different frequencies. This allows for the coherent combining of the multipath components therefore providing a much higher beamforming gain.
\item The possibility to acquire full \gls{csi} from just the demodulation reference signals, without long and costly beam training operations. This enables overhead-free beamforming, as well as the ability to design better and more complex beams. It also makes it possible to efficiently perform localization and sensing tasks by observing the channel estimates, without impacting the communication.
\item The ability to perform spatial multiplexing, i.e., to send \edit{individual data streams} to different receivers at the same time and frequency, separating the \edit{data stream by appropriately combining} the signal from each antenna.
\end{itemize}
Unfortunately, this architecture may not be feasible, especially on the \gls{ue} side, due to cost and power consumption constraints. Additionally, despite in a multi-user environment the resource allocation might be localized in frequency, a \gls{ue} implementing this architecture typically receives on the whole bandwidth at all time with the same capabilities, even though the data meant for it may occupy only a small portion of the entire bandwidth. 

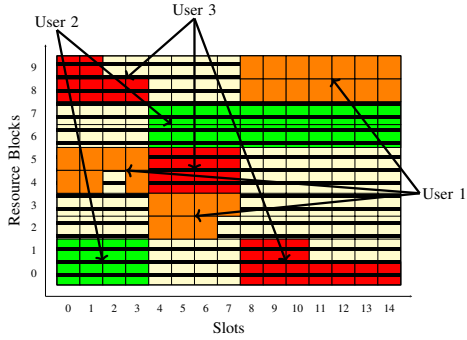
\begin{figure}[t]
\centering
\resizebox{0.7\linewidth}{!}{\begin{tikzpicture}
    \draw [-stealth](-0.5,-0.5) -- (15.5,-0.5);
    \draw [-stealth](-0.5,-0.5) -- (-0.5,10.5);
    \node[font=\huge, anchor=north] at (7.5, -1.5) {Slots};
    \node[rotate=90,font=\huge, anchor=south] at (-1.5, 5) {Resource Blocks};
    \fill[fill=yellow, opacity=0.2] (0,0) -- (15,0) -- (15,10) -- (0, 10) -- cycle; 
    \draw (0,0) grid (15,10);

\foreach \i in {0,...,9}
{
          \node[font=\Large] at (-1, \i + 0.5) {\i};  
}

\foreach \i in {0,...,14}
{
          \node[font=\Large] at (\i + 0.5,-1) {\i};  
}

    \draw[fill=red] (0,8) rectangle ++(1,1);
    \draw[fill=red] (0,9) rectangle ++(1,1);
    \draw[fill=red] (1,8) rectangle ++(1,1);
    \draw[fill=red] (1,9) rectangle ++(1,1);
    \draw[fill=red] (2,8) rectangle ++(1,1);
    \draw[fill=red] (3,8) rectangle ++(1,1);
    \draw[fill=red] (4,4) rectangle ++(1,1);
    \draw[fill=red] (4,5) rectangle ++(1,1);
    \draw[fill=red] (5,4) rectangle ++(1,1);
    \draw[fill=red] (5,5) rectangle ++(1,1);
    \draw[fill=red] (6,4) rectangle ++(1,1);
    \draw[fill=red] (6,5) rectangle ++(1,1);
    \draw[fill=red] (7,4) rectangle ++(1,1);
    \draw[fill=red] (7,5) rectangle ++(1,1);
    \draw[fill=red] (8,0) rectangle ++(1,1);
    \draw[fill=red] (8,1) rectangle ++(1,1);
    \draw[fill=red] (9,0) rectangle ++(1,1);
    \draw[fill=red] (9,1) rectangle ++(1,1);
    \draw[fill=red] (10,0) rectangle ++(1,1);
    \draw[fill=red] (10,1) rectangle ++(1,1);
    \draw[fill=red] (11,0) rectangle ++(1,1);
    \draw[fill=red] (12,0) rectangle ++(1,1);
    \draw[fill=red] (13,0) rectangle ++(1,1);
    \draw[fill=red] (14,0) rectangle ++(1,1);

    \draw[fill=green] (0,0) rectangle ++(1,1);
    \draw[fill=green] (0,1) rectangle ++(1,1);
    \draw[fill=green] (1,0) rectangle ++(1,1);
    \draw[fill=green] (1,1) rectangle ++(1,1);
    \draw[fill=green] (2,0) rectangle ++(1,1);
    \draw[fill=green] (2,1) rectangle ++(1,1);
    \draw[fill=green] (3,0) rectangle ++(1,1);
    \draw[fill=green] (3,1) rectangle ++(1,1);
    \draw[fill=green] (4,6) rectangle ++(1,1);
    \draw[fill=green] (4,7) rectangle ++(1,1);
    \draw[fill=green] (5,6) rectangle ++(1,1);
    \draw[fill=green] (5,7) rectangle ++(1,1);
    \draw[fill=green] (6,6) rectangle ++(1,1);
    \draw[fill=green] (6,7) rectangle ++(1,1);
    \draw[fill=green] (7,6) rectangle ++(1,1);
    \draw[fill=green] (7,7) rectangle ++(1,1);
    \draw[fill=green] (8,6) rectangle ++(1,1);
    \draw[fill=green] (8,7) rectangle ++(1,1);
    \draw[fill=green] (9,6) rectangle ++(1,1);
    \draw[fill=green] (9,7) rectangle ++(1,1);
    \draw[fill=green] (10,6) rectangle ++(1,1);
    \draw[fill=green] (10,7) rectangle ++(1,1);
    \draw[fill=green] (11,6) rectangle ++(1,1);
    \draw[fill=green] (12,6) rectangle ++(1,1);
    \draw[fill=green] (13,6) rectangle ++(1,1);
    \draw[fill=green] (14,6) rectangle ++(1,1);
    \draw[fill=green] (11,7) rectangle ++(1,1);
    \draw[fill=green] (12,7) rectangle ++(1,1);
    \draw[fill=green] (13,7) rectangle ++(1,1);
    \draw[fill=green] (14,7) rectangle ++(1,1);

    \draw[pattern={mylines[size= 5pt,line width=1.5pt,angle=40]},
        pattern color=black] (0,0) rectangle (15,10);

    \draw[fill=orange] (0,5) rectangle ++(1,1);
    \draw[fill=orange] (0,4) rectangle ++(1,1);
    \draw[fill=orange] (1,5) rectangle ++(1,1);
    \draw[fill=orange] (1,4) rectangle ++(1,1);
    \draw[fill=orange] (2,5) rectangle ++(1,1);
    \draw[fill=orange] (3,5) rectangle ++(1,1);
    \draw[fill=orange] (3,4) rectangle ++(1,1);
    \draw[fill=orange] (4,3) rectangle ++(1,1);
    \draw[fill=orange] (4,2) rectangle ++(1,1);
    \draw[fill=orange] (5,3) rectangle ++(1,1);
    \draw[fill=orange] (5,2) rectangle ++(1,1);
    \draw[fill=orange] (6,3) rectangle ++(1,1);
    \draw[fill=orange] (6,2) rectangle ++(1,1);
    \draw[fill=orange] (7,3) rectangle ++(1,1);
    \draw[fill=orange] (8,8) rectangle ++(1,1);
    \draw[fill=orange] (8,9) rectangle ++(1,1);
    \draw[fill=orange] (9,8) rectangle ++(1,1);
    \draw[fill=orange] (9,9) rectangle ++(1,1);
    \draw[fill=orange] (10,8) rectangle ++(1,1);
    \draw[fill=orange] (10,9) rectangle ++(1,1);
    \draw[fill=orange] (11,8) rectangle ++(1,1);
    \draw[fill=orange] (11,9) rectangle ++(1,1);
    \draw[fill=orange] (12,8) rectangle ++(1,1);
    \draw[fill=orange] (12,9) rectangle ++(1,1);
    \draw[fill=orange] (13,8) rectangle ++(1,1);
    \draw[fill=orange] (13,9) rectangle ++(1,1);
    \draw[fill=orange] (14,8) rectangle ++(1,1);
    \draw[fill=orange] (14,9) rectangle ++(1,1);

\node[font=\huge, anchor=east] (u1) at (18,4) {User 1};
\draw[->, line width=1mm] (u1.west) -- (3,5);
\draw[->, line width=1mm] (u1.west) -- (6,3);
\draw[->, line width=1mm] (u1.west) -- (12,9);

\node[font=\huge] (u2) at (0,11.5) {User 2};
\draw[->, line width=1mm] (u2.south) -- (2,1);
\draw[->, line width=1mm] (u2.south) -- (5,7);

\node[font=\huge] (u3) at (6,12) {User 3};
\draw[->, line width=1mm] (u3.south) -- (3,9);
\draw[->, line width=1mm] (u3.south) -- (6,5);
\draw[->, line width=1mm] (u3.south) -- (10,1);

\end{tikzpicture}}
\caption{Example resource grid.}
\label{fig:res_grid}
\end{figure}

For example, let us consider the resource grid depicted in Fig.~\ref{fig:res_grid}.
\noindent Here the horizontal axis represents time, the vertical axis frequency and the squares the \glspl{rb} that can be assigned to users. The \glspl{rb} are according to the user they are allocated to. Let us now consider user $1$ (orange). We can note how, out of the $150$ available \glspl{rb}, it is exploiting only $28$, which corresponds to roughly $18.7\%$. With a classical beamforming system the user would be receiving the signal for all the \glspl{rb}, thus wasting $81.3\%$ of the data acquired. Note that, despite being more expensive and complex, even the fully digital beamforming architecture suffers from this inefficiency. 

To overcome the limitations and inefficiencies discussed in this section, we propose a new architecture described in the following.

\subsection{Proposed architechture} \label{sec:proposed_architechture}

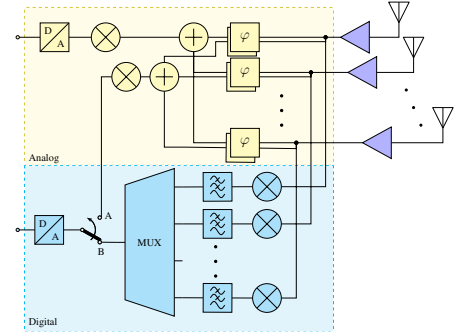
\begin{figure}[t]
\centering
\resizebox{0.7\linewidth}{!}{
%
%
\begin{tikzpicture}

\def \apos {-7}
\def \ashift {0.5}
\def \posend {-13}
\def \shiftapos {\apos + 0.5}
\def \mixpos {-10}
\def \shiftmixpos {\mixpos - 0.5}
\def \dmixpos {-4.5}
\def \ampendpos {-3}
\def \antpos {0}
\def \donepos {-2.5}
\def \dtwopos {-3}
\def \dthreepos {-3.5}
\def \lineone {0}
\def \linetwo {-1.2}
\def \linethree {-3.6}

\def \aone {\lineone + \ashift}
\def \atwo {\linetwo + \ashift}

\def \athree  {\linethree + \ashift}

\def \dlineone {-5.1}
\def \dlinetwo {-6.37}
\def \dlinethree {-8.9}

\def \posouttext {\posend - 0.8}

\def \chaintoshift {0.15}

\tikzset{m1/.style={muxdemux, muxdemux def={Lh=7, Rh=9, NL=1, NB=0, NR=4}}}

\draw


node[adder, fill=yellow!30!white] at (\apos -1,\linetwo - \chaintoshift) (add2) {}

(\donepos,\lineone) to[short, *-] node[short]{} (\donepos,\lineone - \chaintoshift)
to[short] node[short]{} (\donepos-1.95,\lineone - \chaintoshift)
to [phaseshifter, fill=yellow!30!white] node[short]{}(\apos+0.75,\lineone - \chaintoshift)
to[short] node[short]{} (\apos+0.75,\lineone - \chaintoshift - 0.5)
to[short] node[short]{} (\apos-1,\lineone - \chaintoshift - 0.5)
to[short] node[short]{} (add2.north)

(\dtwopos,\linetwo) to[short, *-] node[short]{} (\dtwopos,\linetwo - \chaintoshift)
to[short] node[short]{} (\dtwopos-0.25,\linetwo - \chaintoshift)
to [phaseshifter, fill=yellow!30!white] node[short]{}(add2.east)

(\dthreepos,\linethree) to[short, *-] node[short]{} (\dthreepos,\linethree - \chaintoshift)
to[short] node[short]{} (\dthreepos,\linethree - 0.2)
to [phaseshifter, fill=yellow!30!white] node[short]{} (\apos -0.3,\linethree - \chaintoshift)
to[short] node[short]{} (\apos -1,\linethree - \chaintoshift)
to[short] node[short]{} (add2.south)

node[cute spdt down arrow]  at (-10.5,-6.575) (sw1) {}

node[mixer, fill=yellow!30!white] at (\apos-2.3,\linetwo - \chaintoshift) (mixwb2) {}

 (sw1.out 1) |- (mixwb2.west)
 
 (mixwb2.east) to (add2.west)

node[adder, fill=yellow!30!white] at (\apos,0) (add) {}

node[bareantenna] (ant1) at (\antpos , \aone) {}
(ant1.center)  to node[short]{} (\antpos,  \lineone)
to [amp, fill=blue!30!white] node[short]{}(\ampendpos,\lineone) 
to node[short]{}(-4,\lineone) 
to [phaseshifter, fill=yellow!30!white] (add.east) 

node[bareantenna] (ant2)  at (\antpos + 0.5, \atwo) {}
(ant2.center)  to node[short]{} (\antpos+ 0.5,  \linetwo)
to [amp, fill=blue!30!white] node[short]{}(\ampendpos,\linetwo) 
to node[short]{}(-4,\linetwo) 
to [phaseshifter, fill=yellow!30!white] node[short]{}(\shiftapos,\linetwo)
to node[short]{}( \apos,\linetwo)
 to (add.south)
 
 node[bareantenna] (ant3)  at (\antpos + 1.5,  \athree) {}
(ant3.center)  to node[short]{} (\antpos+ 1.5,  \linethree)
to [amp, fill=blue!30!white] node[short]{}(\ampendpos, \linethree) 
to  node[short]{}(-4, \linethree) 
to [phaseshifter, fill=yellow!30!white] node[short]{}(\shiftapos, \linethree)
to node[short]{}( \apos, \linethree)
 to (add.south)
 
 node[mixer, fill=yellow!30!white] at (\mixpos,0) (mixwb) {}
 (add.west) to (mixwb.east)
 (mixwb.west) to[adc, -o, fill=yellow!30!white] node[short](analogend){}(\posend,0)
 

 
 node[m1, fill=cyan!30!white] (mux1) at (-8.5, -7) {MUX}
 
  node[mixer, fill=cyan!30!white] at (\dmixpos,\dlineone) (mixer1) {}
(\donepos,\lineone) to[short, *-] node[short]{} (\donepos,\dlineone)
to  (mixer1.east)

(mixer1.west) to [lowpass, fill=cyan!30!white] (mux1.rpin 1)

  node[mixer, fill=cyan!30!white] at (\dmixpos,\dlinetwo) (mixer2) {}
(\dtwopos,\linetwo) to[short, *-] node[short]{} (\dtwopos,\dlinetwo)
to  (mixer2.east)

(mixer2.west) to [lowpass, fill=cyan!30!white]  (mux1.rpin 2)

  node[mixer, fill=cyan!30!white] at (\dmixpos,\dlinethree) (mixer3) {}
(\dthreepos, \linethree) to[short, *-] node[short]{} (\dthreepos,\dlinethree)
to  (mixer3.east)

(mixer3.west) to [lowpass, fill=cyan!30!white]  (mux1.rpin 4)

 node[cute spdt down arrow]  at (-10.5,-6.575) (sw1) {}
 
  (sw1.in) to[adc, -o, fill=cyan!30!white] node[short](analogend){}(\posend,-6.575)

(mux1.lpin 1) to[short] (sw1.out 2)
 

 ;

\node[] at (-9.9, -6.1) {A};
\node[] at (-10.15, -7.3) {B};

 \filldraw[color=black, fill=black, thick] (0.25,-1.8) circle (1pt) ;
\filldraw[color=black, fill=black, thick] (0.5,-2.4) circle (1pt) ;
 \filldraw[color=black, fill=black, thick] (0.75,-3) circle (1pt) ;

\filldraw[color=black, fill=black, very thick] (-4,-2) circle (1pt) ;
\filldraw[color=black, fill=black, very thick] (-4,-2.5) circle (1pt) ;
 \filldraw[color=black, fill=black, very thick] (-4,-3) circle (1pt) ;
 
 \filldraw[color=black, fill=black, very thick] (-6.25,-7.15) circle (1pt) ;
\filldraw[color=black, fill=black, very thick] (-6.25,-7.65) circle (1pt) ;
 \filldraw[color=black, fill=black, very thick] (-6.25,-8.15) circle (1pt) ;
 
\node at (\posouttext, \lineone) {};

\begin{scope}[on background layer]
	\draw[cyan!80!white, dashed, line width = 2] (-12.75, -4.4) -- (-2.25, -4.4) -- (-2.25, -10) -- (-12.75, -10) -- 	cycle;
	
	\fill[cyan!10!white] (-12.75, -4.4) -- (-2.25, -4.4) -- (-2.25, -10) -- (-12.75, -10);

	\draw[yellow!80!black, dashed, line width = 2] (-12.75, -4.3) -- (-2.25, -4.3) -- (-2.25,1) -- (-12.75, 1) -- cycle;
	
	\fill[yellow!10!white] (-12.75, -4.3) -- (-2.25, -4.3) -- (-2.25, 1) -- (-12.75, 1);
\end{scope}

  \node[anchor=west] at (-12.75,-9.75) 
    {Digital};

  \node[anchor=west] at (-12.75,-4.15) 
    {Analog};
 
\end{tikzpicture}}
\caption{Proposed hardware architecture. The wideband analog beamforming and the narrowband fully-digital beamforming blocks are highlighted in yellow and blue, respectively.
}
\label{fig:architecture}
\end{figure}

We require a system that is both wideband, to support high data rates, and high rank, \edit{to better exploit the channel diversity}, while maintaining practical costs and power consumption. While a fully digital \gls*{mimo} architecture would satisfy the former constraints, it fails on the latter, i.e., its implementation is inherently costly and power-hungry due to the need for multiple high-speed \glspl{adc}. 
Hence, we propose a new architecture \edit{\cite{arch_patent}}, depicted in Fig.~\ref{fig:architecture}. 

 This architecture comprises $N$ antennas connected via a certain number of classical analog beamforming \gls{rf} chains (marked in yellow in the figure), which operate on the full bandwidth $B_A$ of the system. In particular, the signal from each antenna is amplified by a \gls{lna}, phase shifted and then added together in the analog domain. Subsequently, the combined signal is down-converted and digitized by a single high-speed \gls{adc}. In addition to the analog beamforming system, for each antenna we implement an additional dedicated \gls*{rf} chain (marked in light blue) to implement digital beamforming on a smaller portion of $B_A$. In particular, the signal from each antenna can be extracted after the front-end \glspl{lna}, individually down-converted to baseband and filtered with a \gls*{lpf}.  When the switch in Fig.~\ref{fig:architecture} is in position B, one analog \gls{rf} chain is deactivated and its \gls{adc} is repurposed to digitalize the multiplexing of all the low bandwidth individual antenna signals, in order to obtain a low bandwidth digital beamforming chain. More specifically, since $N$ antennas are multiplexed into a single \gls{adc} that can support a bandwidth $B_A$, the digital beamforming chain operates on a bandwidth $B_D = \frac{B_A}{N}$.
 For the sake of simplicity, in this paper we will focus on the case with 2 \gls{rf} chains, one implementing only analog beamforming, while the other that can be repurposed to implement digital beamforming.
  Furthermore, we note that the mixers belonging to the digital beamforming chain do not necessarily need to be fed with the same \gls{lo} frequency of the analog beamforming, therefore we can assume the center frequency of the digital beamforming part is anywhere within or outside the band of the analog beamforming.  
  Note that, the proposed adaptive architecture can entail any number of RF chains, of which any subset can be converted to digital beamforming, clearly with higher cost, complexity, and power consumption. \edit{Moreover, though the proposed architecture is for the receiver side, a similar architecture can be replicated for a transmitter, thus the results of this study might extend also to uplink transmissions.}


We note that, with this method, we can allocate the samples of the \gls{adc} to acquire a smaller part of the bandwidth with higher rank, thus reducing the number of wasted samples used by classical architectures to digitize the full bandwidth even when data is localized in frequency. \edit{Moreover, this ease of acquisition can enable better and more complete \gls{csi} estimation, which can better capture the rich multipath of industrial environments. This allows for more sophisticated beam design strategies as compared to the classic pencil beam, which can better exploit the channel diversity.}

To summarize, in this paper we assume that the following architecture can operate in two modes
\begin{itemize}
\item \textit{Hybrid mode}, where the switch is in position A, and both \glspl{adc} are connected to an analog beamforming chain and operate on the full bandwidth $B_A$.
\item \textit{Heterogeneous mode}, where the switch is in position B. Here the first \gls{adc} is still connected to the analog bemaforming chain, and operates on the full bandwidth $B_A$, whereas the second \gls{adc} is multiplexed between all the antennas, and therefore operates in digital beamforming on a reduced bandwidth $\frac{B_A}{N}$.
\end{itemize}

Given the proposed architecture, it is apparent that due to the low bandwidth requirements of the digital beamforming \gls{rf} chains, and the re-purposing of \gls{adc} meant for the analog beamforming chain, the complexity, power consumption and costs are not significantly increased. These aspects are better analyzed in Sec.~\ref{sec:arch_compl_cost}.

\section{Complexity, power consumption and cost analysis} \label{sec:arch_compl_cost}
In this section, the power consumption and complexity of the proposed architecture are discussed and compared with those of the classical hybrid beamforming and fully digital beamforming architectures. We consider a system with a $28$GHz carrier and $400$MHz bandwidth, and select off-the-shelf components for the comparison. These were chosen through a thorough web search, filtering for the characteristics required to meet the system specifications. The search result is then sorted by price and the cheapest component is chosen. 

It should be noted that the selected components are just examples to illustrate the main design trade-offs, and they might not entirely reflect the final cost and power consumption of an integrated device. This exercise however can give important insights into the costs and complexity associated with the hardware aspects of the proposed architecture. 

The components selected for the comparison, as well as their main characteristics, are listed in the following:\footnote{All prices refer to those listed in the DigiKey website as of \edit{September 2023}, for a quantity of 25 pieces.}
\begin{itemize}
\item The mixer is the Mini-Circuits MDB-54H+ \cite{mixer_datasheet}. It operates in the $20$-$50$GHz frequency range, and requires an \gls{lo} power of $15$dBm, (roughly $32$mW). \edit{Its price is $30.66$€}.
\item The \gls{adc} is the texas instruments ADS5403 \cite{ADC_datasheet}. It is capable of sampling at a rate of up to $500$Msps with a $12$bit resolution, which is sufficient to handle the required maximum bandwidth of $400$MHz. It has a total power dissipation of $1$W. At the time of writing (May 2023). \edit{Its price is $152.61$€}. 
\item The \gls{lna} is the Mini-Circuits PMA3-313GLN+ \cite{LNA_datasheet}. It operates between $26.5$ and $31$GHz with a gain of $18$dB. It is designed for a $4$V power supply with a biasing current of $78$mA, for a total power consumption of $312$mW. \edit{Its price is $33.43$€}.
\end{itemize}
Note that, since modern communication systems typically use IQ sampling, each \gls{rf} chain requires $2$ mixers and $2$ \glspl{adc}. All other components (multiplexer, phase shifters and filters) are cheap and have low power consumption. \\

Let us now consider a system with $32$ antennas. The required amount of components to build such system is listed in Tab.~\ref{tab:components}.
\begin{table}[t]
\caption{Components required by each type of architecture.}
\centering
\begin{tabular}{c|c|c|c}
\hline
\textbf{Architecture} & \textbf{\# of mixers} & \textbf{\# of \glspl{adc}} & \textbf{\# of \glspl{lna}} \\ \hline
Hybrid  & $4$ & $4$ & \multirow{3}{*}{$32$}\\ 
Proposed  & $64$ & $4$ &\\ 
Fully digital & $64$ & $64$ & \\ 
\end{tabular}
\label{tab:components}
\end{table}
Assuming the \gls{lo} generation has an efficiency of $50\%$, the power needed to generate the reference signal for each mixer is of $2 \times 32$mW = $64$mW. From this, we can compute the power consumption of each system, as listed in  Tab.~\ref{tab:power_consumption}.

\begin{table}[t]
\caption{Comparison of the power consumption of each architecture.}
\centering
\begin{tabular}{c|c|c|c|c}
\hline
\textbf{Architecture} & \textbf{Mixers} & \textbf{\glspl{adc}} & \textbf{\glspl{lna}} & \textbf{Total} \\ \hline
Hybrid  & $256$mW & $4$W & \multirow{3}{*}{$10$W} & $14.26$W\\ 
Proposed  & $4.1$W & $4$W & & $18.1$W\\ 
Fully digital & $4.1$W & $64$W & & $78.1$W \\
\end{tabular}
\label{tab:power_consumption}
\end{table}

It is clear from Tab.~\ref{tab:power_consumption} that the power consumption due to the \glspl{adc} is way more significant than that of the mixers. As a consequence, the power consumption of fully digital architecture, which requires as many \gls{adc} as the number of antennas, is more than $5$ times larger than that of the hybrid beamforming architecture. Instead, the proposed architecture, which can reuse the same ADC for multiple antennas, has only slightly larger power consumption ($+30$\%) with respect to the hybrid architecture. It should be noted that, in industrial applications, \glspl{ue} that need high performance communications typically do not suffer from power constraints. They are in fact typically mounted on robots that consume hundreds or thousands of Watts, making the transceiver power consumption negligible. Instead, the limiting factor is the thermal output of the device, as this needs to be kept within operating temperatures. In the example above, the hybrid and proposed architectures can be cooled with a passive heatsink of limited size, whereas the fully digital architecture is likely to require a large heatsink. The latter might be problematic in terms of space constraints. Additionally, it might result in the inclusion of an active cooling component, such as a fan or a water pump, that can be an issue as they suffer from wear, especially in the harsh conditions of manufacturing plants, and therefore require active maintenance and are prone to cause disruptions when they get damaged. 
 
Another factor to consider is the cost of the components. Clearly, the costs listed \edit{above} are for low quantities, and are not the final production costs. Moreover, such a system would be most likely integrated in a few chips, instead of being composed of individual parts for each components. However, for the sake of this evaluation we will assume that the cost ratios between the components are similar to those of the final implementation. This is  a reasonable assumption for a rough estimate, assuming the dimensional factors of the macro-components are somehow maintained in their integrated version, and considering that the cost of the \glspl{ic} depends on their footprint in silicon. From the number of components indicated in Tab.~\ref{tab:components} we can hence estimate an indicative cost of each architecture, as shown in Tab.~\ref{tab:cost}.

\begin{table}[t]
\caption{Comparison of the cost of each architecture.}
\centering
\begin{tabular}{c|c|c|c|c}
\hline
\textbf{Architechture} & \textbf{Mixers} & \textbf{\glspl{adc}} & \textbf{\glspl{lna}} & \textbf{Total} \\ \hline
Hybrid  & $122.64$€ & $610.44$€ & \multirow{3}{*}{$1069.76$€}  & $1802.84$€\\ 
Proposed  & $1962.24$€ & $610.44$€ & & $3642.44$€\\ 
Fully digital & $1962.24$€ & $9767.04$€ & & $12799.00$€\\
\end{tabular}
\label{tab:cost}
\end{table}

We can observe that the cost of the proposed architecture is about twice that of the classical hybrid beamforming, whereas the fully digital architecture is more than $7$ times more expensive. In future factories, where we expect a huge number of connected devices, such difference might have a very significant \gls{capex} impact. Moreover the relaxed cooling requirements of the proposed architecture will also reduce \gls{opex}. While an in depth estimation of \gls{capex} and \gls{opex} is beyond the scope of this study, it should be clear that the proposed architecture is significantly cheaper than a fully digital beamforming solution.

Finally, we consider the complexity associated with the various architectures. To roughly quantify the complexity of the processing required to use such architectures, we consider the data rate of the samples generated by the systems. In particular, the hybrid beamforming architecture has $4$ \glspl{adc} generating $500$ million $12$-bit samples per second, for a total of $24$Gbps. The proposed architecture, having the same amount of \glspl{adc}, will generate the same amount of data to be processed by the baseband. In contrast, the fully digital beamforming architecture uses $64$ \glspl{adc}, aggregating to a total data rate of $384$Gbps of data. If we assume that the processing time scales linearly with the amount of bits generated by the \glspl{adc}, the fully digital beamforming architecture would consequently require $16$ times more processing power compared to the proposed and hybrid architectures. This would also lead to $16$ times increase in the power consumption in the digital domain for the fully digital beamforming architecture. In addition, we observe that not all of the signal processing algorithms employed in modern receiver have linear complexity, therefore the gap in the required processing capabilities and power consumption will likely be even larger. 

Hence we have shown that in, terms of cost, complexity and power consumption, the proposed architecture is clearly an interesting middle ground between the classical hybrid beamforming and the complex fully digital beamforming architectures. Next, in Sec.~\ref{sec:mimo_sched} we demonstrate the benefits of the proposed architecture in a multi-user resource allocation scenario. 
\section{Multi-user allocation for partial band \gls{mimo}} \label{sec:mimo_sched}
\subsection{Proportional Fairness Resource Allocation}
This section discusses the use of the proposed architecture for communication in a multi-user setting. We assume that the \gls{bs} is equipped with a fully digital (or at least high rank) architecture capable of fully exploiting the rank of the channel. The \glspl{ue} are equipped with the proposed architecture which is also capable of exploiting the full channel rank in the digital beamforming part of the bandwidth.

Specifically, we assume that there are $U$ users with an architecture consisting of $N$ antennas sharing the total bandwidth $B_A$. Recalling the resource allocation example shown in Fig.~\ref{fig:res_grid}, we can see that user $1$ (in orange) would potentially benefit from digital beamforming in the central \glspl{rb} for the first four slots, then in \glspl{rb} $2$ and $3$ for the following 4 slots, and in the top \glspl{rb} for the rest of the time. This is however impractical, as it requires re-configuring the \gls{lo} of the digital chains every few slots. Such re-configuration can take from tens of $\mu$s to some ms \cite{PLL_datasheet}. To avoid this issue, we propose that at the time of connection the \gls{bs} informs the \gls{ue} whether to use the second \gls{rf} chain for digital beamforming and on which part of the band. Subsequently, at the time of \gls{mac} scheduling, the \gls{bs} preferentially schedules each \gls{ue} on the frequencies where it performs digital beamforming.

\begin{figure}[t]
\centering
\resizebox{0.7\linewidth}{!}{ 
\begin{tikzpicture}
    \draw [-stealth](-0.5,-0.5) -- (15.5,-0.5);
    \draw [-stealth](-0.5,-0.5) -- (-0.5,10.5);
\node[,font=\huge, anchor=north] at (7.5, -1.5) {Slots};
    \node[rotate=90,font=\huge, anchor=south] at (-1.5, 5) {Resource Blocks};
    \fill[fill=yellow, opacity=0.2] (0,0) -- (15,0) -- (15,10) -- (0, 10) -- cycle; 
    \draw (0,0) grid (15,10);

\foreach \i in {0,...,9}
{
          \node[font=\Large] at (-1, \i + 0.5) {\i};  
}

\foreach \i in {0,...,14}
{
          \node[font=\Large] at (\i + 0.5,-1) {\i};  
}

    \draw[fill=red] (0,8) rectangle ++(1,1);
    \draw[fill=red] (0,9) rectangle ++(1,1);
    \draw[fill=red] (1,8) rectangle ++(1,1);
    \draw[fill=red] (1,9) rectangle ++(1,1);
    \draw[fill=red] (2,8) rectangle ++(1,1);
    \draw[fill=red] (3,8) rectangle ++(1,1);
    \draw[fill=red] (4,8) rectangle ++(1,1);
    \draw[fill=red] (4,9) rectangle ++(1,1);
    \draw[fill=red] (5,8) rectangle ++(1,1);
    \draw[fill=red] (5,9) rectangle ++(1,1);
    \draw[fill=red] (6,8) rectangle ++(1,1);
    \draw[fill=red] (6,9) rectangle ++(1,1);
    \draw[fill=red] (7,8) rectangle ++(1,1);
    \draw[fill=red] (7,9) rectangle ++(1,1);
    \draw[fill=red] (8,8) rectangle ++(1,1);
    \draw[fill=red] (8,9) rectangle ++(1,1);
    \draw[fill=red] (9,8) rectangle ++(1,1);
    \draw[fill=red] (9,9) rectangle ++(1,1);
    \draw[fill=red] (10,8) rectangle ++(1,1);
    \draw[fill=red] (10,9) rectangle ++(1,1);
    \draw[fill=red] (11,9) rectangle ++(1,1);
    \draw[fill=red] (12,9) rectangle ++(1,1);
    \draw[fill=red] (13,9) rectangle ++(1,1);
    \draw[fill=red] (14,9) rectangle ++(1,1);

    \draw[fill=green] (0,0) rectangle ++(1,1);
    \draw[fill=green] (0,1) rectangle ++(1,1);
    \draw[fill=green] (1,0) rectangle ++(1,1);
    \draw[fill=green] (1,1) rectangle ++(1,1);
    \draw[fill=green] (2,0) rectangle ++(1,1);
    \draw[fill=green] (2,1) rectangle ++(1,1);
    \draw[fill=green] (3,0) rectangle ++(1,1);
    \draw[fill=green] (3,1) rectangle ++(1,1);
    \draw[fill=green] (4,0) rectangle ++(1,1);
    \draw[fill=green] (4,1) rectangle ++(1,1);
    \draw[fill=green] (5,0) rectangle ++(1,1);
    \draw[fill=green] (5,1) rectangle ++(1,1);
    \draw[fill=green] (6,0) rectangle ++(1,1);
    \draw[fill=green] (6,1) rectangle ++(1,1);
    \draw[fill=green] (7,0) rectangle ++(1,1);
    \draw[fill=green] (7,1) rectangle ++(1,1);
    \draw[fill=green] (8,0) rectangle ++(1,1);
    \draw[fill=green] (8,1) rectangle ++(1,1);
    \draw[fill=green] (9,0) rectangle ++(1,1);
    \draw[fill=green] (9,1) rectangle ++(1,1);
    \draw[fill=green] (10,0) rectangle ++(1,1);
    \draw[fill=green] (10,1) rectangle ++(1,1);
    \draw[fill=green] (11,0) rectangle ++(1,1);
    \draw[fill=green] (12,1) rectangle ++(1,1);
    \draw[fill=green] (13,0) rectangle ++(1,1);
    \draw[fill=green] (14,0) rectangle ++(1,1);
    \draw[fill=green] (11,0) rectangle ++(1,1);
    \draw[fill=green] (12,0) rectangle ++(1,1);
    \draw[fill=green] (13,0) rectangle ++(1,1);
    \draw[fill=green] (14,1) rectangle ++(1,1);

   \draw[pattern={mylines[size= 5pt,line width=1.5pt,angle=40]},
        pattern color=black] (0,0) rectangle (15,10);

    \draw[fill=orange] (0,5) rectangle ++(1,1);
    \draw[fill=orange] (0,4) rectangle ++(1,1);
    \draw[fill=orange] (1,5) rectangle ++(1,1);
    \draw[fill=orange] (1,4) rectangle ++(1,1);
    \draw[fill=orange] (2,5) rectangle ++(1,1);
    \draw[fill=orange] (3,5) rectangle ++(1,1);
    \draw[fill=orange] (3,4) rectangle ++(1,1);
    \draw[fill=orange] (4,5) rectangle ++(1,1);
    \draw[fill=orange] (4,4) rectangle ++(1,1);
    \draw[fill=orange] (5,5) rectangle ++(1,1);
    \draw[fill=orange] (5,4) rectangle ++(1,1);
    \draw[fill=orange] (6,5) rectangle ++(1,1);
    \draw[fill=orange] (6,4) rectangle ++(1,1);
    \draw[fill=orange] (7,5) rectangle ++(1,1);
    \draw[fill=orange] (8,5) rectangle ++(1,1);
    \draw[fill=orange] (8,4) rectangle ++(1,1);
    \draw[fill=orange] (9,5) rectangle ++(1,1);
    \draw[fill=orange] (9,4) rectangle ++(1,1);
    \draw[fill=orange] (10,5) rectangle ++(1,1);
    \draw[fill=orange] (10,4) rectangle ++(1,1);
    \draw[fill=orange] (11,5) rectangle ++(1,1);
    \draw[fill=orange] (11,4) rectangle ++(1,1);
    \draw[fill=orange] (12,5) rectangle ++(1,1);
    \draw[fill=orange] (12,4) rectangle ++(1,1);
    \draw[fill=orange] (13,5) rectangle ++(1,1);
    \draw[fill=orange] (13,4) rectangle ++(1,1);
    \draw[fill=orange] (14,5) rectangle ++(1,1);
    \draw[fill=orange] (14,4) rectangle ++(1,1);

\node[font=\huge] (u1) at (17,5) {User 1};
\node[font=\huge] (u1) at (17,1) {User 2};
\node[font=\huge] (u1) at (17,9) {User 3};

\end{tikzpicture}}
\caption{Example resource grid with proposed architecture.}
\label{fig:res_grid_mimo}
\end{figure}
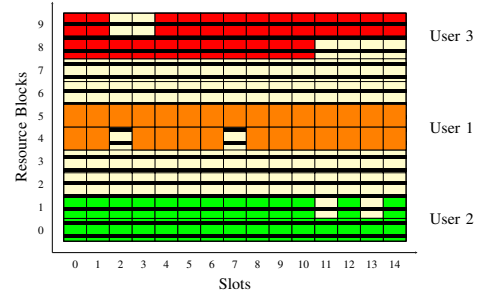

In the example, the \gls{bs} could instruct user $2$ to perform digital beamforming in \glspl{rb} $0$ and $1$, user $1$ on \glspl{rb} $4$ and $5$, and user $3$ on \glspl{rb} $8$ and $9$. It would then schedule the data according to the new resource grid in Fig.~\ref{fig:res_grid_mimo}. In this case, the data would fit entirely within the band where the \glspl{ue} perform digital beamforming. With this allocation, the first \gls{rf} chain, which operates with analog beamforming, still spends $81.3\%$ of its samples to acquire the area covered with the black strips, but the second \gls{rf} chain, operating in digital beamforming, only acquires $30$ \glspl{rb}, out of which $28$ are intended for that user, thus only wasting $6.7\%$ of the \gls{adc} samples. In total, the \textquote{efficiency} of the proposed architecture, intended as the fraction of \gls{adc} samples on frequency bands carrying data, can be estimated as 
\begin{equation}
100 - \frac{81.3+6.7}{2} = 56\%
\end{equation} 
while that of the classical fully connected hybrid beamforming architecture in the same scenario would be $18.7$\%, thus confirming that the proposed architecture is significantly more efficient than the classical one when \glspl{ue}' data are allocated to sub portions of the whole bandwidth using digital beamforming.

\begin{figure}[t]
\centering
         \includegraphics{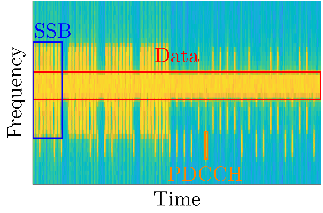}
\caption{Example 5G downlink spectrogram with \gls{ssb} and \gls{pdcch} information located outside the \glspl{rb} designated for digital beamforming.}
\label{fig:controlsigexp}
\end{figure}

\begin{figure}[t]
\centering
\resizebox{0.7\linewidth}{!}{ 
\begin{tikzpicture}
    \draw [-stealth](-0.5,-0.5) -- (15.5,-0.5);
    \draw [-stealth](-0.5,-0.5) -- (-0.5,10.5);
\node[,font=\huge, anchor=north] at (7.5, -1.5) {Slots};
    \node[rotate=90,font=\huge, anchor=south] at (-1.5, 5) {Resource Blocks};
    \fill[fill=yellow, opacity=0.2] (0,0) -- (15,0) -- (15,10) -- (0, 10) -- cycle; 
    \draw (0,0) grid (15,10);

\foreach \i in {0,...,9}
{
          \node[font=\Large] at (-1, \i + 0.5) {\i};  
}

\foreach \i in {0,...,14}
{
          \node[font=\Large] at (\i + 0.5,-1) {\i};  
}

    \draw[fill=red] (0,8) rectangle ++(1,1);
    \draw[fill=red] (0,9) rectangle ++(1,1);
    \draw[fill=red] (1,8) rectangle ++(1,1);
    \draw[fill=red] (1,9) rectangle ++(1,1);
    \draw[fill=red] (2,8) rectangle ++(1,1);
    \draw[fill=red] (3,8) rectangle ++(1,1);
    \draw[fill=red] (4,8) rectangle ++(1,1);
    \draw[fill=red] (4,9) rectangle ++(1,1);
    \draw[fill=red] (5,8) rectangle ++(1,1);
    \draw[fill=red] (5,9) rectangle ++(1,1);
    \draw[fill=red] (6,8) rectangle ++(1,1);
    \draw[fill=red] (6,9) rectangle ++(1,1);
    \draw[fill=red] (7,8) rectangle ++(1,1);
    \draw[fill=red] (7,9) rectangle ++(1,1);
    \draw[fill=red] (8,8) rectangle ++(1,1);
    \draw[fill=red] (8,9) rectangle ++(1,1);
    \draw[fill=red] (9,8) rectangle ++(1,1);
    \draw[fill=red] (9,9) rectangle ++(1,1);
    \draw[fill=red] (10,8) rectangle ++(1,1);
    \draw[fill=red] (10,9) rectangle ++(1,1);
    \draw[fill=red] (11,9) rectangle ++(1,1);
    \draw[fill=red] (12,9) rectangle ++(1,1);
    \draw[fill=red] (13,9) rectangle ++(1,1);
    \draw[fill=red] (14,9) rectangle ++(1,1);

    \draw[fill=green] (0,0) rectangle ++(1,1);
    \draw[fill=green] (0,1) rectangle ++(1,1);
    \draw[fill=green] (1,0) rectangle ++(1,1);
    \draw[fill=green] (1,1) rectangle ++(1,1);
    \draw[fill=green] (2,0) rectangle ++(1,1);
    \draw[fill=green] (2,1) rectangle ++(1,1);
    \draw[fill=green] (3,0) rectangle ++(1,1);
    \draw[fill=green] (3,1) rectangle ++(1,1);
    \draw[fill=green] (4,0) rectangle ++(1,1);
    \draw[fill=green] (4,1) rectangle ++(1,1);
    \draw[fill=green] (5,0) rectangle ++(1,1);
    \draw[fill=green] (5,1) rectangle ++(1,1);
    \draw[fill=green] (6,0) rectangle ++(1,1);
    \draw[fill=green] (6,1) rectangle ++(1,1);
    \draw[fill=green] (7,0) rectangle ++(1,1);
    \draw[fill=green] (7,1) rectangle ++(1,1);
    \draw[fill=green] (8,0) rectangle ++(1,1);
    \draw[fill=green] (8,1) rectangle ++(1,1);
    \draw[fill=green] (9,0) rectangle ++(1,1);
    \draw[fill=green] (9,1) rectangle ++(1,1);
    \draw[fill=green] (10,0) rectangle ++(1,1);
    \draw[fill=green] (10,1) rectangle ++(1,1);
    \draw[fill=green] (11,0) rectangle ++(1,1);
    \draw[fill=green] (12,1) rectangle ++(1,1);
    \draw[fill=green] (13,0) rectangle ++(1,1);
    \draw[fill=green] (14,0) rectangle ++(1,1);
    \draw[fill=green] (11,0) rectangle ++(1,1);
    \draw[fill=green] (12,0) rectangle ++(1,1);
    \draw[fill=green] (13,0) rectangle ++(1,1);
    \draw[fill=green] (14,1) rectangle ++(1,1);

   \draw[pattern={mylines[size= 5pt,line width=1.5pt,angle=40]},
        pattern color=black] (0,0) rectangle (15,10);

    \draw[fill=orange] (0,5) rectangle ++(1,1);
    \draw[fill=orange] (0,4) rectangle ++(1,1);
    \draw[fill=orange] (1,5) rectangle ++(1,1);
    \draw[fill=orange] (1,4) rectangle ++(1,1);
    \draw[fill=orange] (2,5) rectangle ++(1,1);
    \draw[fill=orange] (3,5) rectangle ++(1,1);
    \draw[fill=orange] (3,4) rectangle ++(1,1);
    \draw[fill=orange] (4,5) rectangle ++(1,1);
    \draw[fill=orange] (4,4) rectangle ++(1,1);
    \draw[fill=orange] (5,5) rectangle ++(1,1);
    \draw[fill=orange] (5,4) rectangle ++(1,1);
    \draw[fill=orange] (6,5) rectangle ++(1,1);
    \draw[fill=orange] (6,4) rectangle ++(1,1);
    \draw[fill=orange] (6,3) rectangle ++(1,1);
    \draw[fill=orange] (6,2) rectangle ++(1,1);
    \draw[fill=orange] (7,5) rectangle ++(1,1);
    \draw[fill=orange] (7,4) rectangle ++(1,1);
    \draw[fill=orange] (7,3) rectangle ++(1,1);
    \draw[fill=orange] (7,2) rectangle ++(1,1);
    \draw[fill=orange] (7,6) rectangle ++(1,1);
    \draw[fill=orange] (7,7) rectangle ++(1,1);
    \draw[fill=orange] (8,5) rectangle ++(1,1);
    \draw[fill=orange] (8,4) rectangle ++(1,1);
    \draw[fill=orange] (8,3) rectangle ++(1,1);
    \draw[fill=orange] (8,2) rectangle ++(1,1);
    \draw[fill=orange] (8,6) rectangle ++(1,1);
    \draw[fill=orange] (8,7) rectangle ++(1,1);
    \draw[fill=orange] (9,5) rectangle ++(1,1);
    \draw[fill=orange] (9,4) rectangle ++(1,1);
    \draw[fill=orange] (10,5) rectangle ++(1,1);
    \draw[fill=orange] (10,4) rectangle ++(1,1);
    \draw[fill=orange] (11,5) rectangle ++(1,1);
    \draw[fill=orange] (11,4) rectangle ++(1,1);
    \draw[fill=orange] (12,5) rectangle ++(1,1);
    \draw[fill=orange] (12,4) rectangle ++(1,1);
    \draw[fill=orange] (13,5) rectangle ++(1,1);
    \draw[fill=orange] (13,4) rectangle ++(1,1);
    \draw[fill=orange] (14,5) rectangle ++(1,1);
    \draw[fill=orange] (14,4) rectangle ++(1,1);

\node[font=\huge] (u1) at (17,5) {User 1};
\node[font=\huge] (u1) at (17,1) {User 2};
\node[font=\huge] (u1) at (17,9) {User 3};

\end{tikzpicture}}
\caption{Example resource grid with proposed architecture, where the data does not fit entirely within the \glspl{rb} defined for digital beamforming.}
\label{fig:res_grid_mimo_excess}
\end{figure}
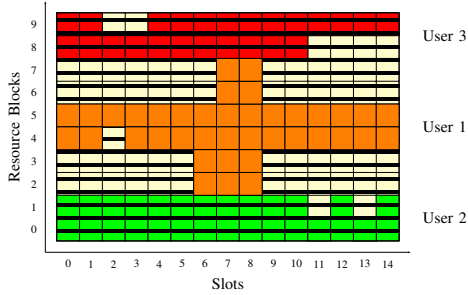

Note that the analog \gls{rf} chain is still necessary to decode the resource block allocation. Practically, there is control information sent by the \gls{bs} that will reside outside the digital beamforming \glspl{rb}. In order to decode this information, the analog \gls{rf} will be needed. This is illustrated in Fig.~\ref{fig:controlsigexp}, which shows an example of spectrogram of a 5G downlink signal with a single user. Here it can be observed that some control signals, such as the \gls{ssb} and \gls{pdcch}, resides outside the \glspl{rb} used by the \gls{ue} for receiving data. Moreover, the analog \gls{rf} chain will also be needed in case the data do not fit within the \glspl{rb} used for digital beamforming, and have to be partially allocated to other \glspl{rb}. This case is illustrated in Fig.~\ref{fig:res_grid_mimo_excess}. 


Next, to analyze the performance of the proposed architecture from the aspect of spectral efficiency, let us assume that all the users are equal, and experience the following:
\begin{itemize}
\item A \gls{se} of $C_A$ bit/s/Hz when using the analog \gls{rf} chain in \textit{heterogeneous mode}.
\item A \gls{se} of $C_D$ bit/s/Hz when using the digital \gls{rf} chain in \textit{heterogeneous mode}.
\item A \gls{se} of $C_H$ bit/s/Hz when using the \textit{hybrid mode}.
\end{itemize}

It is important to state that, in typical situations, we have:
\begin{equation}
C_A \leq C_H \leq C_D. \label{eq:spectral_eff_order}
\end{equation} 

\noindent Assuming that the assignment of the digital beamforming sub-band is performed with minimum overlap, since each user can perform digital beamforming on $\frac{1}{N} B_A$, the fraction of bandwidth that has at least one user with digital beamforming on it is given by: 
\begin{equation}
\zeta_D = \min \left( 1, \frac{U}{N} \right). \label{eq:fraction_MIMO}
\end{equation}
If all \glspl{ue} are configured to perform digital beamforming in a part of the band, the maximum average \gls{se} achievable with digital beamforming enabled can be expressed as:
\begin{equation}
C^{(D)}_{max} = (1-\zeta_D) C_A + \zeta_D C_D = C_A + \zeta_D (C_D-C_A).
\end{equation}
Comparing this to the \gls{se} of the classical hybrid beamforming system, we observe that
\begin{equation}
 C^{(D)}_{max} < C_H  \iff \zeta_D (C_D-C_A) < C_H - C_A.
\end{equation}
Recalling \eqref{eq:spectral_eff_order}, the term $C_D - C_A$ is always positive, therefore we have
\begin{equation}
 C^{(D)}_{max} < C_H  \iff \zeta_D  <\frac{C_H - C_A}{C_D-C_A}.  \label{eq:maxzeta}
\end{equation}
\noindent Moreover, from \eqref{eq:spectral_eff_order} we also conclude that $(C_D-C_A) \geq (C_H - C_A)$, therefore

\begin{equation}
0 < \frac{C_H - C_A}{C_D-C_A} \leq 1,
\end{equation}
\noindent which assures the resulting value of $\zeta_D$ is meaningful.
We can now replace the definition of $\zeta_D$ in \eqref{eq:maxzeta} to obtain:

\begin{equation}
\frac{U}{N} < \frac{C_H - C_A}{C_D-C_A} \Leftrightarrow U < \frac{N(C_H - C_A)}{C_D-C_A}.
\end{equation}

Under this condition, it is always better to utilize the two \gls{rf} chains for hybrid beamforming. This also allows us to write the overall maximum achievable average \gls{se} as a function of the number of users, which is given as:
\begin{equation}
C_{max} = \begin{cases}
C_H & \text{if} \: U < \frac{N(C_H - C_A)}{C_D-C_A}; \\
C_A + \frac{U}{N} (C_D-C_A) & \text{if}    \: \frac{N(C_H - C_A)}{C_D-C_A} < U <N;\\
C_D & \text{otherwise}.
\end{cases} \label{eq:cmax}
\end{equation}

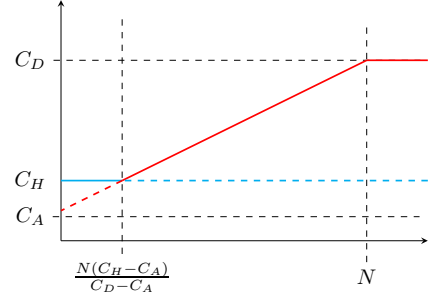
\begin{figure}[t]
\centering
\resizebox{\fwidth}{\fheight}{\def \posca{0.4}
\def \posch{1}
\def \poscd{3}

\def \poscmin{1}
\def \poscmax{5}

\begin{tikzpicture}
\draw[-stealth] (0,0) -- (6,0);
\draw[-stealth] (0,0) -- (0,4);

\node(ca) at(-0.5,\posca) {$C_A$};
\node(ch)  at(-0.5,\posch) {$C_H$};
\node(cd)  at(-0.5,\poscd) {$C_D$};

\draw[dashed] (ca.east) -- +(6,0);
\draw[dashed] (cd.east) -- +(6,0);

\node(umin) at(\poscmin, -0.6) {$\frac{N(C_H - C_A)}{C_D-C_A}$};
\node(umax) at(\poscmax, -0.6) {$N$};

\draw[dashed] (umin.north) -- +(0,4);
\draw[dashed] (umax.north) -- +(0,4);

\draw[thick, cyan] (0,\posch)--(\poscmin,\posch);
\draw[thick, red] (\poscmin,\posch) -- (\poscmax,\poscd) ;
\draw[thick, red]  (\poscmax,\poscd) -- (6,\poscd) ;

\draw[thick, red, dashed] (\poscmin,\posch) -- +(-1,-0.5) ;

\draw[thick, cyan,dashed] (6,\posch)--(\poscmin,\posch);

\end{tikzpicture}}
\caption{Maximum achievable \gls{se} as a function of the number of user.}
\label{fig:max_rate_mimo}
\end{figure}

This function is illustrated in Fig.~\ref{fig:max_rate_mimo}. The cyan line represents the maximum  \gls{se} achievable when operating in \textit{hybrid mode}, whereas the red line represents the maximum performance of the \textit{heterogeneous mode}. The solid line represents the value of $C_{max}$, which is the maximum of the two.

From the figure, we can clearly identify the three regions corresponding to the cases in \eqref{eq:cmax}. In the first part, the \gls{se} of the \textit{hybrid mode} dominates, because there are not enough users to have digital beamforming on a significant portion of the bandwidth when operating in \textit{heterogeneous mode}. In the central region, the number of users is sufficient for the combination of analog and digital beamforming of the \textit{heterogeneous mode} to achieve a better \gls{se} than the \textit{hybrid mode}. However, the sub-channels used for digital beamforming are not enough to cover the whole available bandwidth, so that $C_{max}< C_{D}$. Finally, in the third region, all portions of the band are used by at least one \gls{ue} with digital beamforming, therefore potentially achieving fully digital \gls{mimo} across the whole bandwidth. These considerations refer to the maximum achievable rate. However, there is no guarantee that the actual system performance will be close to that bound. In particular, the bound is achievable only if all \glspl{rb} that can be used for digital beamforming are actually allocated to the users that were given those channels to perform digital beamforming. 

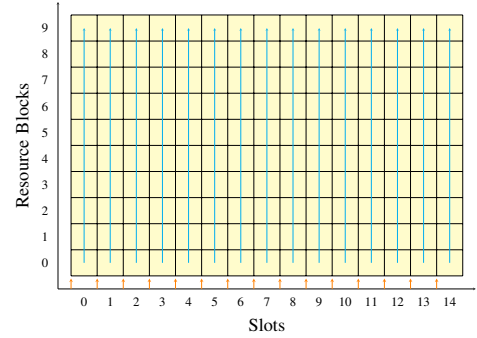
\begin{figure}[t]
\centering
\resizebox{0.7\linewidth}{!}{ 
\begin{tikzpicture}
    \draw [-stealth](-0.5,-0.5) -- (15.5,-0.5);
    \draw [-stealth](-0.5,-0.5) -- (-0.5,10.5);
    \node[,font=\huge, anchor=north] at (7.5, -1.5) {Slots};
    \node[rotate=90,font=\huge, anchor=south] at (-1.5, 5) {Resource Blocks};
    \fill[fill=yellow, opacity=0.2] (0,0) -- (15,0) -- (15,10) -- (0, 10) -- cycle; 
    \draw (0,0) grid (15,10);

\foreach \i in {0,...,9}
{
          \node[font=\Large] at (-1, \i + 0.5) {\i};  
}

\foreach \i in {0,...,14}
{
        \node[font=\Large] at (\i + 0.5,-1) {\i};  
        \draw [-stealth, mycolor1] (\i +0.5,0.5) -- (\i +0.5, 9.5);
        \draw [-stealth, mycolor2, thick] (\i ,-0.5) -- (\i , -0.1);

}

\end{tikzpicture}
 }
\caption{Allocation process.}
\label{fig:alloc_proc}
\end{figure}

 To verify that it is possible to exploit such capability in a more realistic scenario, we consider a system implementing \gls{ofdma} with $R$ \glspl{rb}, and receivers with $N$ antennas. For the sake of simplicity, we assume that a \gls{rb} lasts $1$s and has a bandwidth of $1$Hz, thus the capacity of a \gls{rb} is equal to the spectral efficiency. When operating in \textit{heterogeneous mode}, the second \gls{rf} chain of user $u \in \{1,...,U\}$ performs digital beamforming in  $\alpha = \left\lfloor\frac{R}{N}\right\rfloor$ \glspl{rb}, and the allocations to different users are either orthogonal or completely overlapping. In particular, user $u$ performs digital beamforming on \glspl{rb} $r \in \mathcal{D}_u$, where 
\begin{align}
\mathcal{D}_u =  & \big\{ \left(u\alpha\bmod R \right),\left(u\alpha + 1\bmod R \right), \nonumber\\&...,\left(u\alpha+\alpha-1 \bmod R \right) \big\}.
\end{align}
\noindent We also define the set of users performing digital beamforming in \gls{rb} $r$ as
\begin{equation}
\mathcal{U}_r = \left\{u : r \in \mathcal{D}_u  \right\}.
\end{equation}
 For each user, we keep an estimate of its average rate $C_u$. After each slot, the estimate is updated to $C_u = \gamma C_u + (1-\gamma) \bar{C}_u$, where $\bar{C}_u$ is the total rate experienced by that user in the slot, that under the considered assumptions, above the amount of data transferred in that slot, and $\gamma \in \{0,1\}$.
The allocation process is depicted in Fig.~\ref{fig:alloc_proc}. We assume the arrivals, depicted in the figure as orange arrows, happen between time slots. At each slot, we iterate over the \glspl{rb} in the order shown by the cyan arrows and assign each \gls{rb} $r$ to the \gls{ue} $u$ that satisfies the following constraints:
\begin{itemize}
\item Its buffer is not empty.
\item It has the  highest \gls{pf} weight $W = \frac{C_{r}}{C_u}$.
\end{itemize}
In \textit{hybrid mode} we consider $C_{r} = C_H$, whereas for the \textit{heterogeneous mode} we consider 
\begin{equation}
C_{r} = \begin{cases}
C_D, & \text{if} \: r \in \mathcal{D}_u,\\
C_A, & \text{otherwise}.
\end{cases}
\end{equation}
If multiple \glspl{ue} satisfy such constraint, we choose one at random.  When an \gls{rb} is assigned to a \gls{ue}, the \gls{ue}'s queue is decreased by $C_{r}$.

We assume each node $n$, independently of the others, generates a random number $v_n$ of bytes at every slot. Let  $F_v(a) = \Pr[v_n \leq a]$ be the \gls{cdf} of $v_n$, identical for all nodes. 
Now, let us focus on the set $\mathcal{D}_u$ of \glspl{rb} assigned to certain user $u$.
Let $V_u = \sum_{n \in \mathcal{U}_r} v_n$ be the aggregate traffic generated by all the nodes that are pre-assigned at the \glspl{rb} in $\mathcal{D}_u$. Let $F_{V_u}(\cdot)$ be the \gls{cdf} of $V_u$, which can be computed from $F_v(\cdot)$ with standard methods. For the sake of analysis, we also assume that:
\begin{enumerate}[label={A\arabic*}]
\item At the beginning of the slot, all nodes in $\mathcal{U}_r$ have empty queues. \label{ass:empty}
\item Their average throughput is such that the \gls{pf} algorithm will preferably assign the \glspl{rb} in $\mathcal{U}_r$ to these nodes, before considering other nodes which will use the \gls{rb} with analog beamforming. \label{ass:allocation}
\end{enumerate}

Let $X_u$ be the number of \glspl{rb} in $D_u$ assigned to users that can perform digital beamforming. We hence have that 
\begin{equation}
X_u \leq \min\left( \alpha, \left\lceil \frac{V_u}{C_D} \right\rceil \right),
\label{eq:X}
\end{equation}
and its expectation can be computed by deriving the \gls{pdf} of $V_u$ from the finite difference of its \gls{cdf} and averaging over all possible values $h$ of $X_u$ up to $\alpha$, to obtain 
\begin{align}
\E[X_u] = &\sum_{h=1}^{\alpha-1} h(F_{V_u}(hC_d)- F_{V_u}((h-1)C_D)) \nonumber \\&+ \alpha (1-F_{V_u}((\alpha-1) C_D)).
\end{align}

An estimate of the total system throughput can hence be obtained as
\begin{equation}
\tilde{C}(U,F_v) = \min \left( G , \sum_{u=1}^{N} \E[X_u]C_D +(\alpha-\E[X_u])C_A \right)
\end{equation}
\noindent where $G$ is the overall traffic generated by the $U$ nodes in a slot, $u > U \Rightarrow E[X_u] = 0$, i.e. users that are not in the system do not generate traffic, and the summation is up to $N$ because a user $u'>N$ will share the sub-band with user $ u = u' \bmod N$.

\subsection{Results}

In this section, we present the results for the system discussed so far with $R=640$ \glspl{rb} and $N=32$ antennas. The digital beamforming \gls{se} is $C_D = 4$ and the analog beaforming \gls{se} is $C_A = 1$. When operating in \textit{hybrid mode}, the \gls{se} is $C_H=1.5$. Further, each user has a buffer of $1000$bits, and at each slot it generates a random number of bits between $0$ and $2 \Lambda$, where $\Lambda$ is the average generation rate.
With this assumption, $F_{V_{u}}(a)$ is the \gls{cdf} of an Irwin-Hall distribution of parameter $n = \vert D_r \vert$ computed in $\frac{a}{2 \Lambda}$.

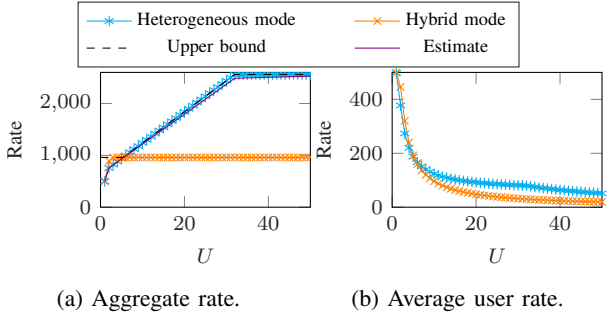
\begin{figure}[t]
     \centering
     \begin{subfigure}[t]{\linewidth}
     	\centering
         \begin{tikzpicture}

\begin{axis}[
    width=0,
    height=0,
    at={(0,0)},
    scale only axis,
    xmin=0,
    xmax=0,
    xtick={},
    ymin=0,
    ymax=0,
    ytick={},
    axis background/.style={fill=white},
    legend style={legend cell align=center, align=center, draw=white!15!black, font=\scriptsize, at={(0, 0)}, anchor=center, /tikz/every even column/.append style={column sep=2em}},
    legend columns=2,
]
\addplot [color=mycolor1, mark=asterisk, mark options={solid, mycolor1}]
  table{%
 0 1
};
\addlegendentry{Heterogeneous mode}

\addplot [color=mycolor2, mark=x, mark options={solid, mycolor2}]
  table{%
 0 1
};
\addlegendentry{Hybrid mode}

\addplot [color=black, dashed]
  table{%
 0 1
};
\addlegendentry{Upper bound}

\addplot [color=mycolor3]
  table{%
 0 1
};
\addlegendentry{Estimate}

\end{axis}

\end{tikzpicture}
     \end{subfigure}
     \begin{subfigure}[t]{0.45\linewidth}
%
%

\begin{tikzpicture}

\begin{axis}[%
width=\fwidth,
height=\fheight,
scale only axis,
xmin=0,
xmax=50,
style={font=\footnotesize\color{white!15!black}},
xlabel={$U$},
xlabel near ticks,
ymin=0,
ymax=2600,
ylabel style={font=\footnotesize\color{white!15!black}},
ylabel={Rate},
ylabel near ticks,
axis background/.style={fill=white},
legend style={at={(0.98,0.02)}, anchor=south east, legend cell align=left, align=left, draw=white!15!black}]

\addplot [color=mycolor1, mark=asterisk, mark options={solid, mycolor1}, clip mode=individual, forget plot]
  table[row sep=crcr]{%
1	497.919\\
2	754.95\\
3	819.829\\
4	879.94\\
5	939.88\\
6	999.88\\
7	1059.88\\
8	1119.907\\
9	1179.301\\
10	1239.937\\
11	1299.73\\
12	1359.94\\
13	1419.772\\
14	1479.91\\
15	1539.703\\
16	1599.904\\
17	1659.94\\
18	1719.928\\
19	1779.919\\
20	1839.925\\
21	1898.53\\
22	1960\\
23	2019.955\\
24	2079.805\\
25	2139.799\\
26	2199.73\\
27	2260\\
28	2318.353\\
29	2378.149\\
30	2439.904\\
31	2499.904\\
32	2559.634\\
33	2558.14\\
34	2558.185\\
35	2558.224\\
36	2558.278\\
37	2558.38\\
38	2558.395\\
39	2558.479\\
40	2558.461\\
41	2558.614\\
42	2558.656\\
43	2558.644\\
44	2558.818\\
45	2558.743\\
46	2558.92\\
47	2558.98\\
48	2558.977\\
49	2559.04\\
50	2559.103\\
};
\addplot [color=mycolor2, mark=x, mark options={solid, mycolor2},clip mode=individual, forget plot]
  table[row sep=crcr]{%
1	499.5045\\
2	892.5795\\
3	960\\
4	960\\
5	960\\
6	960\\
7	960\\
8	960\\
9	960\\
10	960\\
11	960\\
12	960\\
13	960\\
14	960\\
15	960\\
16	960\\
17	960\\
18	960\\
19	960\\
20	960\\
21	960\\
22	960\\
23	960\\
24	960\\
25	960\\
26	960\\
27	960\\
28	960\\
29	960\\
30	960\\
31	960\\
32	960\\
33	960\\
34	960\\
35	960\\
36	960\\
37	960\\
38	960\\
39	960\\
40	960\\
41	960\\
42	960\\
43	960\\
44	960\\
45	960\\
46	960\\
47	960\\
48	960\\
49	960\\
50	960\\
};

\addplot [color=black, dashed, forget plot]
  table[row sep=crcr]{%
0	960\\
5.33333333333333	960\\
32	2560\\
50	2560\\
};

\addplot [color=mycolor3, forget plot]
  table[row sep=crcr]{%
1	500\\
2	755.44\\
3	813.16\\
4	870.88\\
5	928.6\\
6	986.32\\
7	1044.04\\
8	1101.76\\
9	1159.48\\
10	1217.2\\
11	1274.92\\
12	1332.64\\
13	1390.36\\
14	1448.08\\
15	1505.8\\
16	1563.52\\
17	1621.24\\
18	1678.96\\
19	1736.68\\
20	1794.4\\
21	1852.12\\
22	1909.84\\
23	1967.56\\
24	2025.28\\
25	2083\\
26	2140.72\\
27	2198.44\\
28	2256.16\\
29	2313.88\\
30	2371.6\\
31	2429.32\\
32	2487.04\\
33	2489.26072\\
34	2491.48144\\
35	2493.70216\\
36	2495.92288\\
37	2498.1436\\
38	2500.36432\\
39	2502.58504\\
40	2504.80576\\
41	2507.02648\\
42	2509.2472\\
43	2511.46792\\
44	2513.68864\\
45	2515.90936\\
46	2518.13008\\
47	2520.3508\\
48	2522.57152\\
49	2524.79224\\
50	2527.01296\\
};
\end{axis}

\end{tikzpicture}%
         \caption{Aggregate rate.}
         \label{fig:agg_rate_500}
     \end{subfigure}
     \begin{subfigure}[t]{0.45\linewidth}       
%
%
\begin{tikzpicture}

\begin{axis}[%
width=\fwidth,
height=\fheight,
scale only axis,
xmin=0,
xmax=50,
style={font=\footnotesize\color{white!15!black}},
xlabel={$U$},
xlabel near ticks,
ymin=0,
ymax=500,
ylabel style={font=\footnotesize\color{white!15!black}},
ylabel={Rate},
ylabel near ticks,
axis background/.style={fill=white},
legend style={at={(0.98,0.02)}, anchor=south east, legend cell align=left, align=left, draw=white!15!black}]
\addplot [color=mycolor1, mark=asterisk, mark options={solid, mycolor1}, forget plot]
  table[row sep=crcr]{%
1	497.919\\
2	377.475\\
3	273.276333333333\\
4	219.985\\
5	187.976\\
6	166.646666666667\\
7	151.411428571429\\
8	139.988375\\
9	131.033444444444\\
10	123.9937\\
11	118.157272727273\\
12	113.328333333333\\
13	109.213230769231\\
14	105.707857142857\\
15	102.646866666667\\
16	99.994\\
17	97.6435294117647\\
18	95.5515555555556\\
19	93.6799473684211\\
20	91.99625\\
21	90.4061904761905\\
22	89.0909090909091\\
23	87.8241304347826\\
24	86.6585416666667\\
25	85.59196\\
26	84.605\\
27	83.7037037037037\\
28	82.7983214285714\\
29	82.0051379310345\\
30	81.3301333333333\\
31	80.642064516129\\
32	79.9885625\\
33	77.5193939393939\\
34	75.2407352941176\\
35	73.0921142857143\\
36	71.0632777777778\\
37	69.1454054054054\\
38	67.3261842105263\\
39	65.6020256410256\\
40	63.961525\\
41	62.4052195121951\\
42	60.920380952381\\
43	59.5033488372093\\
44	58.1549545454546\\
45	56.8609555555556\\
46	55.6286956521739\\
47	54.4463829787234\\
48	53.3120208333333\\
49	52.225306122449\\
50	51.18206\\
};
\addplot [color=mycolor2, mark=x, mark options={solid, mycolor2}, forget plot]
  table[row sep=crcr]{%
1	499.5045\\
2	446.28975\\
3	320\\
4	240\\
5	192\\
6	160\\
7	137.142857142857\\
8	120\\
9	106.666666666667\\
10	96\\
11	87.2727272727273\\
12	80\\
13	73.8461538461538\\
14	68.5714285714286\\
15	64\\
16	60\\
17	56.4705882352941\\
18	53.3333333333333\\
19	50.5263157894737\\
20	48\\
21	45.7142857142857\\
22	43.6363636363636\\
23	41.7391304347826\\
24	40\\
25	38.4\\
26	36.9230769230769\\
27	35.5555555555556\\
28	34.2857142857143\\
29	33.1034482758621\\
30	32\\
31	30.9677419354839\\
32	30\\
33	29.0909090909091\\
34	28.2352941176471\\
35	27.4285714285714\\
36	26.6666666666667\\
37	25.9459459459459\\
38	25.2631578947368\\
39	24.6153846153846\\
40	24\\
41	23.4146341463415\\
42	22.8571428571429\\
43	22.3255813953488\\
44	21.8181818181818\\
45	21.3333333333333\\
46	20.8695652173913\\
47	20.4255319148936\\
48	20\\
49	19.5918367346939\\
50	19.2\\
};
\end{axis}

\end{tikzpicture}%
         \caption{Average user rate.}
         \label{fig:user_rate_500}
     \end{subfigure}
        \caption{Rate achieved by the user and system for $\Lambda = 500$bps.}
        
        \label{fig:rate_500}
\end{figure}

Fig.~\ref{fig:rate_500} shows the performance achieved with a data generation rate of $\Lambda=500$bps. This rate is sufficient to saturate the total capacity with a few users. In particular, Fig.~\ref{fig:agg_rate_500} shows the aggregate rate of the base station as a function of the number of \glspl{ue}, as well as the maximum achievable rate $C_{max}$ and the estimated rate $\tilde{C}(U, F_v)$. Here we can observe that below $5$ users the \textit{hybrid mode} performs better than the \textit{heterogeneous mode}. We note that, in this case, we have $\frac{N(C_H - C_A)}{C_D-C_A} = 5.33$, therefore the limit of $5$ \glspl{ue} corresponds to what is predicted by the upper bound. We can also observe that after this point the digital beamforming curve tightly follows the bound, confirming that a \gls{pf} scheduler can fully exploit the capabilities of the proposed architecture.\\ Fig.~\ref{fig:user_rate_500} shows the average rate observed by each user. Here we can again observe that the \textit{hybrid mode} performs better only with less than $5$ users, and the \textit{heterogeneous mode} provides a much higher rate for a large number of users.

In Fig.~\ref{fig:rate_500}, we can also observe how the proposed estimate is close to the real rate for a loaded system. This is expected as, for a large traffic, $X_u$ is likely to be close to $\alpha$, and therefore it is expected that most \glspl{rb} are allocated to digital beamforming.

\begin{figure}[t]
     \centering
     \begin{subfigure}[t]{\linewidth}
     	\centering
         \begin{tikzpicture}

\begin{axis}[
    width=0,
    height=0,
    at={(0,0)},
    scale only axis,
    xmin=0,
    xmax=0,
    xtick={},
    ymin=0,
    ymax=0,
    ytick={},
    axis background/.style={fill=white},
    legend style={legend cell align=center, align=center, draw=white!15!black, font=\scriptsize, at={(0, 0)}, anchor=center, /tikz/every even column/.append style={column sep=2em}},
    legend columns=2,
]
\addplot [color=mycolor1, mark=asterisk, mark options={solid, mycolor1}]
  table{%
 0 1
};
\addlegendentry{Heterogeneous mode}

\addplot [color=mycolor2, mark=x, mark options={solid, mycolor2}]
  table{%
 0 1
};
\addlegendentry{Hybrid mode}

\addplot [color=black, dashed]
  table{%
 0 1
};
\addlegendentry{Upper bound}

\addplot [color=mycolor3]
  table{%
 0 1
};
\addlegendentry{Estimate}

\end{axis}

\end{tikzpicture}
     \end{subfigure}
     \begin{subfigure}[t]{0.45\linewidth}
%
%

%
\begin{tikzpicture}

\begin{axis}[%
width=\fwidth,
height=\fheight,
scale only axis,
xmin=0,
xmax=50,
style={font=\footnotesize\color{white!15!black}},
xlabel={$U$},
xlabel near ticks,
ymin=0,
ymax=2600,
ylabel style={font=\footnotesize\color{white!15!black}},
ylabel={Rate},
ylabel near ticks,
axis background/.style={fill=white},
legend style={at={(0.98,0.02)}, anchor=south east, legend cell align=left, align=left, draw=white!15!black}]

\addplot [color=mycolor1, mark=asterisk, mark options={solid, mycolor1}, clip mode=individual, forget plot]
  table[row sep=crcr]{%
1	98.122\\
2	196.834\\
3	302.523\\
4	409.368\\
5	503.423\\
6	609.383\\
7	705.134\\
8	809.332\\
9	914.743\\
10	1016.871\\
11	1107.509\\
12	1208.799\\
13	1317.039\\
14	1407.106\\
15	1497.901\\
16	1576.273\\
17	1651.192\\
18	1715.023\\
19	1774.126\\
20	1836.187\\
21	1897.504\\
22	1957.585\\
23	2017.939\\
24	2078.908\\
25	2137.351\\
26	2197.723\\
27	2258.341\\
28	2318.26\\
29	2377.654\\
30	2438.23\\
31	2498.644\\
32	2557.732\\
33	2558.407\\
34	2557.741\\
35	2558.461\\
36	2558.11\\
37	2557.738\\
38	2558.149\\
39	2557.843\\
40	2557.981\\
41	2558.221\\
42	2557.888\\
43	2558.503\\
44	2558.38\\
45	2558.332\\
46	2558.194\\
47	2558.155\\
48	2558.446\\
49	2558.158\\
50	2558.44\\
};
\addplot [color=mycolor2, mark=x, mark options={solid, mycolor2}, clip mode=individual, forget plot]
  table[row sep=crcr]{%
1	97.3755\\
2	201.8355\\
3	303.6825\\
4	402.747\\
5	505.443\\
6	604.413\\
7	709.83\\
8	812.4135\\
9	914.2605\\
10	958.848\\
11	960\\
12	960\\
13	960\\
14	960\\
15	960\\
16	960\\
17	960\\
18	960\\
19	960\\
20	960\\
21	960\\
22	960\\
23	960\\
24	960\\
25	960\\
26	960\\
27	960\\
28	960\\
29	960\\
30	960\\
31	960\\
32	960\\
33	960\\
34	960\\
35	960\\
36	960\\
37	960\\
38	960\\
39	960\\
40	960\\
41	960\\
42	960\\
43	960\\
44	960\\
45	960\\
46	960\\
47	960\\
48	960\\
49	960\\
50	960\\
};

\addplot [color=black, dashed, forget plot]
  table[row sep=crcr]{%
0	960\\
5.33333333333333	960\\
32	2560\\
50	2560\\
};

\addplot [color=mycolor3, forget plot]
  table[row sep=crcr]{%
1	100\\
2	200\\
3	300\\
4	400\\
5	500\\
6	600\\
7	700\\
8	800\\
9	900\\
10	1000\\
11	1100\\
12	1200\\
13	1271.8\\
14	1320.4\\
15	1369\\
16	1417.6\\
17	1466.2\\
18	1514.8\\
19	1563.4\\
20	1612\\
21	1660.6\\
22	1709.2\\
23	1757.8\\
24	1806.4\\
25	1855\\
26	1903.6\\
27	1952.2\\
28	2000.8\\
29	2049.4\\
30	2098\\
31	2146.6\\
32	2195.2\\
33	2205.118\\
34	2215.036\\
35	2224.954\\
36	2234.872\\
37	2244.79\\
38	2254.708\\
39	2264.626\\
40	2274.544\\
41	2284.462\\
42	2294.38\\
43	2304.298\\
44	2314.216\\
45	2324.134\\
46	2334.052\\
47	2343.97\\
48	2353.888\\
49	2363.806\\
50	2373.724\\
};

\end{axis}

\end{tikzpicture}%
         \caption{Aggregate rate.}
         \label{fig:agg_rate_100}
     \end{subfigure}
     \begin{subfigure}[t]{0.45\linewidth}       
%
%
\begin{tikzpicture}

\begin{axis}[%
width=\fwidth,
height=\fheight,
scale only axis,
xmin=0,
xmax=50,
style={font=\footnotesize\color{white!15!black}},
xlabel={$U$},
xlabel near ticks,
ymin=0,
ymax=120,
ylabel style={font=\footnotesize\color{white!15!black}},
ylabel={Rate},
ylabel near ticks,
axis background/.style={fill=white},
legend style={at={(0.98,0.02)}, anchor=south east, legend cell align=left, align=left, draw=white!15!black}]
\addplot [color=mycolor1, mark=asterisk, mark options={solid, mycolor1}, forget plot]
  table[row sep=crcr]{%
1	98.122\\
2	98.417\\
3	100.841\\
4	102.342\\
5	100.6846\\
6	101.563833333333\\
7	100.733428571429\\
8	101.1665\\
9	101.638111111111\\
10	101.6871\\
11	100.682636363636\\
12	100.73325\\
13	101.310692307692\\
14	100.507571428571\\
15	99.8600666666667\\
16	98.5170625\\
17	97.1289411764706\\
18	95.2790555555555\\
19	93.375052631579\\
20	91.80935\\
21	90.3573333333333\\
22	88.9811363636364\\
23	87.7364782608696\\
24	86.6211666666667\\
25	85.49404\\
26	84.5278076923077\\
27	83.6422592592593\\
28	82.795\\
29	81.9880689655172\\
30	81.2743333333333\\
31	80.6014193548387\\
32	79.929125\\
33	77.5274848484848\\
34	75.2276764705882\\
35	73.0988857142857\\
36	71.0586111111111\\
37	69.128054054054\\
38	67.3197105263158\\
39	65.5857179487179\\
40	63.949525\\
41	62.3956341463415\\
42	60.9020952380952\\
43	59.5000697674419\\
44	58.145\\
45	56.8518222222222\\
46	55.6129130434783\\
47	54.428829787234\\
48	53.3009583333333\\
49	52.207306122449\\
50	51.1688\\
};
\addplot [color=mycolor2, mark=x, mark options={solid, mycolor2}, forget plot]
  table[row sep=crcr]{%
1	97.3755\\
2	100.91775\\
3	101.2275\\
4	100.68675\\
5	101.0886\\
6	100.7355\\
7	101.404285714286\\
8	101.5516875\\
9	101.5845\\
10	95.8848\\
11	87.2727272727273\\
12	80\\
13	73.8461538461538\\
14	68.5714285714286\\
15	64\\
16	60\\
17	56.4705882352941\\
18	53.3333333333333\\
19	50.5263157894737\\
20	48\\
21	45.7142857142857\\
22	43.6363636363636\\
23	41.7391304347826\\
24	40\\
25	38.4\\
26	36.9230769230769\\
27	35.5555555555556\\
28	34.2857142857143\\
29	33.1034482758621\\
30	32\\
31	30.9677419354839\\
32	30\\
33	29.0909090909091\\
34	28.2352941176471\\
35	27.4285714285714\\
36	26.6666666666667\\
37	25.9459459459459\\
38	25.2631578947368\\
39	24.6153846153846\\
40	24\\
41	23.4146341463415\\
42	22.8571428571429\\
43	22.3255813953488\\
44	21.8181818181818\\
45	21.3333333333333\\
46	20.8695652173913\\
47	20.4255319148936\\
48	20\\
49	19.5918367346939\\
50	19.2\\
};
\end{axis}

\end{tikzpicture}%
         \caption{Average user rate.}
         \label{fig:user_rate_100}
     \end{subfigure}
        \caption{Rate achieved by the user and system for $\Lambda = 100$bps.}
        
        \label{fig:rate_100}
\end{figure}
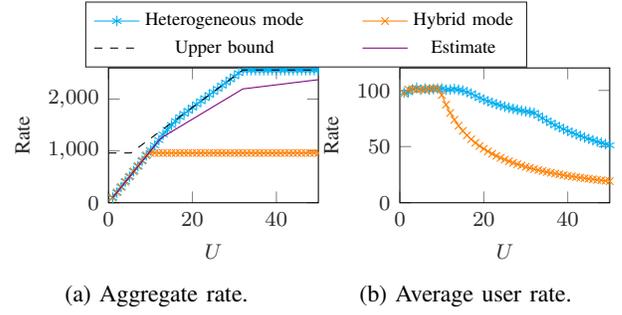

Next, Fig.~\ref{fig:rate_100} shows the results for an application rate of $\Lambda=100$bps. In Fig.~\ref{fig:agg_rate_100} we can see that, in \textit{hybrid mode}, for $U \leq 9$ the rate is limited by the application generation rate. In \textit{heterogeneous mode} instead, this happens for $U \leq 17$, meaning that the proposed scheme is able to support almost twice the number of users at the full rate. In Fig.~\ref{fig:user_rate_100} we can also observe that the average rate observed by each user drops more rapidly for the hybrid beamforming scheme. Thus, the hybrid beamforming system can support only $10$ \glspl{ue} with $90$\% of the required rate, whereas the proposed scheme can support up to $21$ \glspl{ue} for the same required rate. In this case, the estimate turns out to be conservative compared to the actual rate. This is due to the assumption of empty queue (\ref{ass:empty}), which in this case is unrealistic.

\begin{figure}[t]
     \centering
     \begin{subfigure}[t]{\linewidth}
     	\centering
         \begin{tikzpicture}

\begin{axis}[
    width=0,
    height=0,
    at={(0,0)},
    scale only axis,
    xmin=0,
    xmax=0,
    xtick={},
    ymin=0,
    ymax=0,
    ytick={},
    axis background/.style={fill=white},
    legend style={legend cell align=center, align=center, draw=white!15!black, font=\scriptsize, at={(0, 0)}, anchor=center, /tikz/every even column/.append style={column sep=2em}},
    legend columns=2,
]
\addplot [color=mycolor1, mark=asterisk, mark options={solid, mycolor1}]
  table{%
 0 1
};
\addlegendentry{Heterogeneous mode}

\addplot [color=mycolor2, mark=x, mark options={solid, mycolor2}]
  table{%
 0 1
};
\addlegendentry{Hybrid mode}

\addplot [color=black, dashed]
  table{%
 0 1
};
\addlegendentry{Upper bound}

\addplot [color=mycolor3]
  table{%
 0 1
};
\addlegendentry{Estimate}

\end{axis}

\end{tikzpicture}
     \end{subfigure}
     \begin{subfigure}[t]{0.45\linewidth}
%
%
%
\begin{tikzpicture}

\begin{axis}[%
width=\fwidth,
height=\fheight,
scale only axis,
xmin=0,
xmax=50,
style={font=\footnotesize\color{white!15!black}},
xlabel={$U$},
xlabel near ticks,
ymin=0,
ymax=2600,
ylabel style={font=\footnotesize\color{white!15!black}},
ylabel={Rate},
ylabel near ticks,
axis background/.style={fill=white},
legend style={at={(0.98,0.02)}, anchor=south east, legend cell align=left, align=left, draw=white!15!black}]
\addplot [color=mycolor1, mark=asterisk, mark options={solid, mycolor1}, clip mode=individual, forget plot]
  table[row sep=crcr]{%
1	51.07\\
2	104.298\\
3	156.116\\
4	207.9\\
5	264.911\\
6	309.996\\
7	359.257\\
8	414.894\\
9	464.205\\
10	519.149\\
11	567.161\\
12	616.599\\
13	669.299\\
14	726.945\\
15	774.993\\
16	826.315\\
17	872.98\\
18	925.019\\
19	982.917\\
20	1025.65\\
21	1076.028\\
22	1126.185\\
23	1179.178\\
24	1233.439\\
25	1285.25\\
26	1343.842\\
27	1387.603\\
28	1445.505\\
29	1502.853\\
30	1549.999\\
31	1601.508\\
32	1644.957\\
33	1695.274\\
34	1753.753\\
35	1796.674\\
36	1848.388\\
37	1891.579\\
38	1940.053\\
39	1964.29\\
40	2000.665\\
41	2033.569\\
42	2064.022\\
43	2091.247\\
44	2114.167\\
45	2142.457\\
46	2168.383\\
47	2188.402\\
48	2208.625\\
49	2232.376\\
50	2254.393\\
};

\addplot [color=mycolor2, mark=x, mark options={solid, mycolor2}, clip mode=individual, forget plot]
  table[row sep=crcr]{%
1	51.7335\\
2	101.418\\
3	153.09\\
4	205.5615\\
5	255.789\\
6	310.869\\
7	354.726\\
8	405.1845\\
9	463.671\\
10	508.167\\
11	561.1815\\
12	613.7115\\
13	665.9175\\
14	709.0365\\
15	766.368\\
16	810.8475\\
17	863.3355\\
18	914.883\\
19	959.6145\\
20	960\\
21	960\\
22	959.9505\\
23	960\\
24	960\\
25	960\\
26	960\\
27	960\\
28	960\\
29	960\\
30	960\\
31	960\\
32	960\\
33	960\\
34	960\\
35	960\\
36	960\\
37	960\\
38	960\\
39	960\\
40	960\\
41	960\\
42	960\\
43	960\\
44	960\\
45	960\\
46	960\\
47	960\\
48	960\\
49	960\\
50	960\\
};

\addplot [color=black, dashed, forget plot]
  table[row sep=crcr]{%
0	960\\
5.33333333333333	960\\
32	2560\\
50	2560\\
};

\addplot [color=mycolor3, forget plot]
  table[row sep=crcr]{%
1	50\\
2	100\\
3	150\\
4	200\\
5	250\\
6	300\\
7	350\\
8	400\\
9	450\\
10	500\\
11	550\\
12	600\\
13	650\\
14	700\\
15	750\\
16	800\\
17	850\\
18	900\\
19	950\\
20	1000\\
21	1050\\
22	1100\\
23	1150\\
24	1200\\
25	1250\\
26	1300\\
27	1350\\
28	1400\\
29	1450\\
30	1500\\
31	1550\\
32	1600\\
33	1650\\
34	1700\\
35	1750\\
36	1800\\
37	1850\\
38	1900\\
39	1948.504\\
40	1965.376\\
41	1982.248\\
42	1999.12\\
43	2015.992\\
44	2032.864\\
45	2049.736\\
46	2066.608\\
47	2083.48\\
48	2100.352\\
49	2117.224\\
50	2134.096\\
};

\end{axis}

\end{tikzpicture}%
         \caption{Aggregate rate.}
         \label{fig:agg_rate_50}
     \end{subfigure}
     \begin{subfigure}[t]{0.45\linewidth}       
%
%
\begin{tikzpicture}

\begin{axis}[%
width=\fwidth,
height=\fheight,
scale only axis,
xmin=0,
xmax=50,
style={font=\footnotesize\color{white!15!black}},
xlabel={$U$},
xlabel near ticks,
ymin=0,
ymax=55,
ylabel style={font=\footnotesize\color{white!15!black}},
ylabel={Rate},
ylabel near ticks,
axis background/.style={fill=white},
legend style={at={(0.98,0.02)}, anchor=south east, legend cell align=left, align=left, draw=white!15!black}]
\addplot [color=mycolor1, mark=asterisk, mark options={solid, mycolor1}, forget plot]
  table[row sep=crcr]{%
1	51.07\\
2	52.149\\
3	52.0386666666667\\
4	51.975\\
5	52.9822\\
6	51.666\\
7	51.3224285714286\\
8	51.86175\\
9	51.5783333333333\\
10	51.9149\\
11	51.5600909090909\\
12	51.38325\\
13	51.4845384615385\\
14	51.9246428571429\\
15	51.6662\\
16	51.6446875\\
17	51.3517647058824\\
18	51.3899444444444\\
19	51.7324736842105\\
20	51.2825\\
21	51.2394285714286\\
22	51.1902272727273\\
23	51.2686086956522\\
24	51.3932916666667\\
25	51.41\\
26	51.6862307692308\\
27	51.3927037037037\\
28	51.6251785714286\\
29	51.8225172413793\\
30	51.6666333333333\\
31	51.6615483870968\\
32	51.40490625\\
33	51.3719393939394\\
34	51.5809705882353\\
35	51.3335428571429\\
36	51.3441111111111\\
37	51.1237567567568\\
38	51.0540263157895\\
39	50.3664102564103\\
40	50.016625\\
41	49.599243902439\\
42	49.143380952381\\
43	48.6336511627907\\
44	48.04925\\
45	47.6101555555556\\
46	47.1387608695652\\
47	46.5617446808511\\
48	46.0130208333333\\
49	45.558693877551\\
50	45.08786\\
};

\addplot [color=mycolor2, mark=x, mark options={solid, mycolor2}, forget plot]
  table[row sep=crcr]{%
1	51.7335\\
2	50.709\\
3	51.03\\
4	51.390375\\
5	51.1578\\
6	51.8115\\
7	50.6751428571429\\
8	50.6480625\\
9	51.519\\
10	50.8167\\
11	51.0165\\
12	51.142625\\
13	51.2244230769231\\
14	50.6454642857143\\
15	51.0912\\
16	50.67796875\\
17	50.7844411764706\\
18	50.8268333333333\\
19	50.5060263157895\\
20	48\\
21	45.7142857142857\\
22	43.6341136363636\\
23	41.7391304347826\\
24	40\\
25	38.4\\
26	36.9230769230769\\
27	35.5555555555556\\
28	34.2857142857143\\
29	33.1034482758621\\
30	32\\
31	30.9677419354839\\
32	30\\
33	29.0909090909091\\
34	28.2352941176471\\
35	27.4285714285714\\
36	26.6666666666667\\
37	25.9459459459459\\
38	25.2631578947368\\
39	24.6153846153846\\
40	24\\
41	23.4146341463415\\
42	22.8571428571429\\
43	22.3255813953488\\
44	21.8181818181818\\
45	21.3333333333333\\
46	20.8695652173913\\
47	20.4255319148936\\
48	20\\
49	19.5918367346939\\
50	19.2\\
};

\end{axis}

\end{tikzpicture}%
         \caption{Average user rate.}
         \label{fig:user_rate_50}
     \end{subfigure}
        \caption{Rate achieved by the user and system for $\Lambda = 50$bps.}
        
        \label{fig:rate_50}
\end{figure}
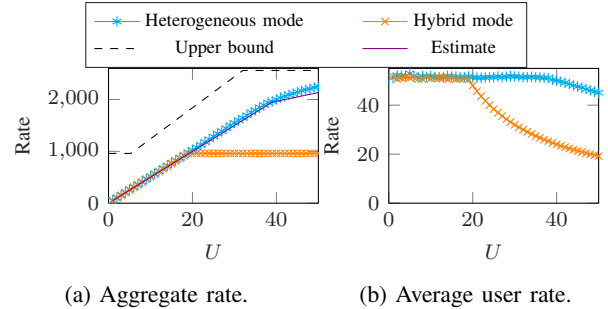

Further, in Fig.~\ref{fig:rate_50} we can observe the results for an application rate of $\Lambda=50$bps. From Fig.~\ref{fig:agg_rate_50} it can be observed that the maximum capacity $C_{max}$ achievable with \textit{heterogeneous mode} is not saturated even at $50$ \glspl{ue}, whereas the hybrid beamforming capacity is saturated at $U=19$. Despite not saturating $C_{max}$, in Fig.~\ref{fig:user_rate_50} we can observe that for $U \geq 38$ the system is unable to support the aggregate generation rate rate of the \glspl{ue}. \edit{This suggests that in this case, the scheduler is unable to fully exploit the digital beamforming, and end up allocating some \glspl{rb} to analog beamforming}.
In this situation the estimate is again close to the simulated value, as the assumption \ref{ass:empty} is more realistic due to the low traffic generated by the individual \glspl{ue}.

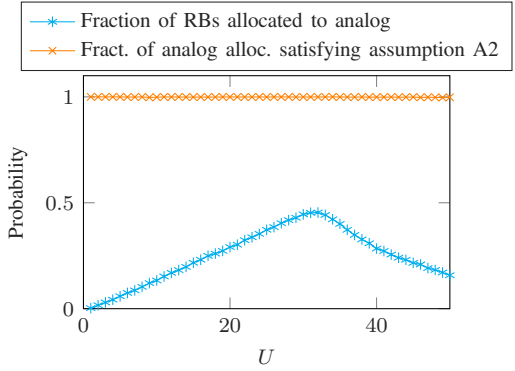
\begin{figure}[t]
\centering
%
%
\begin{tikzpicture}

\begin{axis}[%
width=\singfwidth,
height=\singfheight,
scale only axis,
xmin=0,
xmax=50,
style={font=\footnotesize\color{white!15!black}},
xlabel={$U$},
xlabel near ticks,
ymin=0,
ymax=1.1,
ylabel style={font=\footnotesize\color{white!15!black}},
ylabel={Probability},
ylabel near ticks,
axis background/.style={fill=white},
legend style={at={(0.5,1.02)}, anchor=south, legend cell align=left, align=left, draw=white!15!black}]
\addplot [color=mycolor1, mark=asterisk, mark options={solid, mycolor1}]
  table[row sep=crcr]{%
1	0.002815625\\
2	0.0163453125\\
3	0.0294921875\\
4	0.042553125\\
5	0.05883125\\
6	0.074371875\\
7	0.08654375\\
8	0.1030984375\\
9	0.120809375\\
10	0.1333109375\\
11	0.1522296875\\
12	0.1690765625\\
13	0.1821265625\\
14	0.197365625\\
15	0.2169296875\\
16	0.2311234375\\
17	0.249678125\\
18	0.260090625\\
19	0.2728546875\\
20	0.2908359375\\
21	0.302146875\\
22	0.3234484375\\
23	0.337246875\\
24	0.3532359375\\
25	0.3724828125\\
26	0.3852796875\\
27	0.403790625\\
28	0.4181890625\\
29	0.4294734375\\
30	0.4454421875\\
31	0.4528921875\\
32	0.4540453125\\
33	0.442396875\\
34	0.422003125\\
35	0.4002109375\\
36	0.372346875\\
37	0.348290625\\
38	0.3271734375\\
39	0.3087265625\\
40	0.283940625\\
41	0.2722671875\\
42	0.2555046875\\
43	0.240671875\\
44	0.230796875\\
45	0.2162875\\
46	0.20895\\
47	0.191696875\\
48	0.1830109375\\
49	0.170871875\\
50	0.157971875\\
};

\addlegendentry{Fraction of \glspl{rb} allocated to analog};

\addplot [color=mycolor2, mark=x, mark options={solid, mycolor2}]
  table[row sep=crcr]{%
1	1\\
2	1\\
3	1\\
4	1\\
5	1\\
6	0.999159628555822\\
7	0.999277821910883\\
8	1\\
9	0.997387413021547\\
10	0.998593513754263\\
11	0.999671548954602\\
12	0.999630344980547\\
13	0.999828416022512\\
14	0.998978735532087\\
15	0.999142867432564\\
16	0.999053536056896\\
17	0.999142646156927\\
18	0.999158947001646\\
19	0.99977094034714\\
20	0.999527224863674\\
21	0.99978280430668\\
22	0.999304371349761\\
23	0.999393063316006\\
24	0.999823064435509\\
25	0.999265905725516\\
26	0.999574172983101\\
27	0.999810390595374\\
28	0.999760873707691\\
29	0.999465188111896\\
30	0.999386143684471\\
31	0.999585994183218\\
32	0.999693725502342\\
33	0.999586768102736\\
34	1\\
35	0.999523688679798\\
36	0.999034838146554\\
37	0.998932285357954\\
38	0.999608388135116\\
39	0.99880051623352\\
40	0.998915926525132\\
41	0.998754669987547\\
42	0.999345657797374\\
43	0.998526261117964\\
44	0.998950646537134\\
45	0.999241461018321\\
46	0.997547260110074\\
47	0.99838612392612\\
48	0.997728960871533\\
49	0.996781213994404\\
50	0.998110818777077\\
};
\addlegendentry{Fract. of analog alloc. satisfying assumption \ref{ass:allocation}};

\end{axis}

\end{tikzpicture}%
\caption{Fraction of \glspl{rb} allocated to analog beamforming and reason of the allocation for $\Lambda=50$bps.}
\label{fig:analog_reason}
\end{figure}

Fig.~\ref{fig:analog_reason} shows in cyan the fraction of \glspl{rb} allocated to users performing analog beamforming. It also shows in orange the percentage of such allocations due to the lack of digital beamforming \gls{ue} with data rather than  higher \gls{pf} weight of the analog beamforming \gls{ue}. Indeed, we can see that $100$\% of the analog allocations are done because there is no data from the digital beamforming users. We recall that this corresponds to assumption \ref{ass:allocation}, thus validating the proposed theoretical framework.

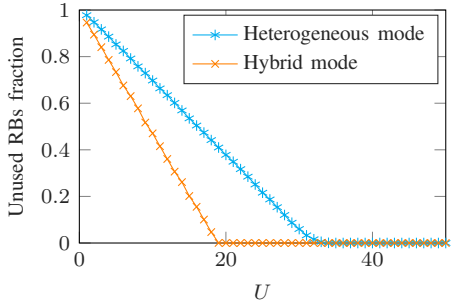
\begin{figure}[t]
\centering
%
%
\begin{tikzpicture}

\begin{axis}[%
width=\singfwidth,
height=\singfheight,
scale only axis,
xmin=0,
xmax=50,
style={font=\footnotesize\color{white!15!black}},
xlabel={$U$},
xlabel near ticks,
ymin=0,
ymax=1,
ylabel style={font=\footnotesize\color{white!15!black}},
ylabel={Unused \glspl{rb} fraction},
ylabel near ticks,
axis background/.style={fill=white},
legend style={at={(0.98,0.98)}, anchor=north east, legend cell align=left, align=left, draw=white!15!black}]
\addplot [color=mycolor1, mark=asterisk, mark options={solid, mycolor1}]
  table[row sep=crcr]{%
1	0.97779375\\
2	0.9475609375\\
3	0.9169484375\\
4	0.8858546875\\
5	0.8524015625\\
6	0.8244765625\\
7	0.7919765625\\
8	0.7580875\\
9	0.7292359375\\
10	0.6973578125\\
11	0.663584375\\
12	0.6345625\\
13	0.601334375\\
14	0.56899375\\
15	0.5349390625\\
16	0.5042015625\\
17	0.4739578125\\
18	0.4417484375\\
19	0.40984375\\
20	0.3782625\\
21	0.3488546875\\
22	0.317096875\\
23	0.2847328125\\
24	0.2511609375\\
25	0.215890625\\
26	0.18688125\\
27	0.1537875\\
28	0.1202359375\\
29	0.087015625\\
30	0.060028125\\
31	0.032809375\\
32	0.016121875\\
33	0.005128125\\
34	0.000778125\\
35	0\\
36	0\\
37	0\\
38	0\\
39	0\\
40	0\\
41	0\\
42	0\\
43	0\\
44	0\\
45	0\\
46	0\\
47	0\\
48	0\\
49	0\\
50	0\\
};
\addlegendentry{Heterogeneous mode};
\addplot [color=mycolor2, mark=x, mark options={solid, mycolor2}]
  table[row sep=crcr]{%
1	0.9461109375\\
2	0.89435625\\
3	0.84053125\\
4	0.7858734375\\
5	0.733553125\\
6	0.676178125\\
7	0.63049375\\
8	0.5779328125\\
9	0.517009375\\
10	0.470659375\\
11	0.4154359375\\
12	0.3607171875\\
13	0.3063359375\\
14	0.2614203125\\
15	0.2017\\
16	0.1553671875\\
17	0.1006921875\\
18	0.046996875\\
19	0.0004015625\\
20	0\\
21	0\\
22	5.15625e-05\\
23	0\\
24	0\\
25	0\\
26	0\\
27	0\\
28	0\\
29	0\\
30	0\\
31	0\\
32	0\\
33	0\\
34	0\\
35	0\\
36	0\\
37	0\\
38	0\\
39	0\\
40	0\\
41	0\\
42	0\\
43	0\\
44	0\\
45	0\\
46	0\\
47	0\\
48	0\\
49	0\\
50	0\\
};
\addlegendentry{Hybrid mode};

\end{axis}

\end{tikzpicture}%
\caption{Fraction of unused \glspl{rb} for $\Lambda=50$bps.}
\label{fig:emptyrbs}
\end{figure}

Lastly, Fig.~\ref{fig:emptyrbs} shows the average fraction of unused \glspl{rb} as a function of $U$ for $\Lambda=50$bps. Here we can observe that for $U \geq 19$ the hybrid beamforming system needs to use all the available \glspl{rb}. This is consistent with the fact that we observe an average rate drop for the \glspl{ue} after that point. In contrast, at $U=19$ the proposed method uses only $59\%$ of the \glspl{rb}, thus saving $41\%$ of the transmission resources. Assuming a constant power spectral density across the allocated \glspl{rb}, this can directly result in a saving in the transmission power.

\section{3GPP signaling for the proposed architecture}
\label{sec:3gpp}

In this section, we analyze the signaling requirements for realizing end-to-end system consisting of the proposed architecture. We also present any potential modifications of the current \gls{3gpp} standard that are needed to utilize the proposed architecture. 
Specifically, we need to be able to perform the following tasks:
\begin{taskenum}
\item Notify the \gls{bs} of the \glspl{ue}' capabilities of trading some analog chains for digital beamforming, as well as the number of antennas it can use in the digital beamforming chain. \label{enum:signaling_cap}
\item Notify the \glspl{ue}' of which configuration should be used based on the status of the network (e.g., number of users and traffic pattern). \label{enum:signaling_conf}
\item Update the \glspl{ue}' configurations as the network conditions change. \label{enum:signaling_reconf}
\item Notify the resource allocation information for utilizing different capabilities in different parts of the band. \label{enum:resourcalloc}
\end{taskenum}

\subsection{Capabilities reporting}
Let us now consider Task~\ref{enum:signaling_cap}. In \gls{3gpp}, \gls{ue} capabilities are reported in the \gls{rrc} \cite{TS_UECap} message, and in particular in a specific \gls{ie} called \textit{UE-NR-Capability}. 
In particular, the information about the \gls{rf} capabilities of the \gls{ue} is contained in the the \textit{rf-Parameters} \gls{ie}
and more specifically in the fields related to the \gls{mimo} and beamforming capabilities are located in the \textit{mimo-ParametersPerBand} \gls{ie} inside the \textit{BandNR} \gls{ie}. 

To advertise the \gls{ue} capability of performing the rank-bandwidth tradeoff for some \gls{rf} chains we introduce a new \gls{ie} called \textit{analogdigitalbfcap} in the \textit{mimo-ParametersPerBand} \gls{ie}, as described in \msgname\ref{asn:MIMO-ParametersPerBand_new}. The newly added \gls{ie} is described in \msgname\ref{asn:ABCap}.

\begin{lstlisting}[float=t,style=3gppmsg,caption={Proposed change to the MIMO-ParametersPerBand \gls{ie}.},label=asn:MIMO-ParametersPerBand_new]
MIMO-ParametersPerBand ::= SEQUENCE {
	...,
	analogdigitalbfcap       SEQUENCE OF ABCap	OPTIONAL
}
\end{lstlisting} 

\begin{lstlisting}[float=t,style=3gppmsg,caption={Proposed structure of the new ABCap \gls{ie}.},label=asn:ABCap]
ABCap :: = SEQUENCE {
	numfixanalogchains    ENUMERATED, 
	numtradablechains ENUMERATED,   
	tradcapab   BIT STRING (Size(8))
}
\end{lstlisting}

\begin{lstlisting}[float=t,style=3gppmsg,caption={Proposed change to the BWP-Downlink \gls{ie}.},label=asn:BWP-Downlink_new]
BWP-Downlink ::=    SEQUENCE {
    bwp-Id              BWP-Id,
    bwp-Common          BWP-DownlinkCommon  
    bwp-Dedicated       BWP-DownlinkDedicated   
    bwp-Tradchain		BWP-Tradchain
    ...
}
\end{lstlisting}

\begin{lstlisting}[float=t,style=3gppmsg,caption={Proposed structure of the BWP-Tradchain \gls{ie}.},label=asn:BWP-Tradchain]
BWP-Tradchain ::=      SEQUENCE {
    genericParameters   BWP, 
    tradcapuse  Bitstring(8) 
    numtradechains Integer(1-maxTradechains)
} 
\end{lstlisting} 

The fields in the \textit{ABCap} \gls{ie} have the following meaning:
\begin{itemize}
\item \textit{numfixanalogchains} is the number of \gls{rf} chains that are exclusively capable of performing analog beamforming.
\item \textit{numtradablechains} is the number of tradable \gls{rf} chains. Concretely, these are the \gls{rf} chains that are capable of performing analog beamforming on the full bandwidth or digital beamforming on a part of the bandwidth.
\item \textit{tradcapab} contains the information on the rank-bandiwdth tradeoff that such \gls{rf} chains can perform. In particular, if the $b$-th bit is set to $1$ the \gls{ue} can use the tradable \gls{rf} chains to acquire the signal from $2^{b+1}$ antennas with bandwidth $\frac{B_A}{2^{b+1}}$. 
\end{itemize}

 \begin{table*}[t]
    \centering
    \caption{Additional parameters to DCI \textit{format1\textunderscore1}}
    \begin{tabular}{M{0.45\textwidth}|c|M{0.45\textwidth}}
    \hline
\textbf{Field}  & \textbf{bits} &  \textbf{Specification} \\ \hline

Modulation and coding scheme for digital beamforming & 5 & Defined in \cite{TS_Alloc} tables 5.1.3.1-1 and 5.1.3.1-2 \\

Antenna port(s) and number of layers for digital beamforming & 4,5,6 & Determined by DMRS Configuration Specified in \cite{TS_DCI} tables 7.3.1.2.2-1 to 7.3.1.2.2-4 \\

    \end{tabular}
    \label{tab:DCI_1_1_add}
\end{table*} 

\subsection{\gls{ue} configuration}
We now investigate the signaling to perform Tasks~\ref{enum:signaling_conf} and \ref{enum:signaling_reconf}. These tasks are again realized via the \gls{rrc} message \cite{TS_RRC}, specifically in an \gls{ie} named \textit{ServingCellConfig}.
This message contains the list \textit{downlinkBWP-ToAddModList} which contains a set of \gls{bwp} configurations, of the format \textit{BWP-Downlink} 
to be used by the \gls{ue}. To configure the \gls{ue} to perform digital beamforming in a specific part of the band, we propose to change this message as described in \msgname\ref{asn:BWP-Downlink_new}. In particular, we add the \gls{ie} \textit{bwp-Tradchain}, which is described in \msgname\ref{asn:BWP-Tradchain}.

The new fields added in the \textit{BWP-Tradchain} format are the following:
 \begin{itemize}
 \item \textit{genericParameters} is the field describing the location and bandwidth on which digital beamforming should be performed. It is of the standard format \textit{BWP}.
 \item \textit{tradcapuse} informs the \gls{ue} of what tradeoff should be used, and follows the same format of the \textit{tradcapab} field in the newly proposed \textit{ABCap} \gls{ie}.
 \item \textit{numtradechains} informs the \gls{ue} of how many of the tradable \gls{rf} chains need to be configured to perform digital beamforming. 
 \end{itemize}

\subsection{Resource allocation}
 The resource allocation information (Task~\ref{enum:resourcalloc}) to be provided to the \glspl{ue} is specified in the \gls{dci} message \cite{TS_DCI}. In particular, the downlink data allocation is specified in the \gls{dci} \textit{format1\textunderscore1}. 
 In this message, the \textit{Frequency domain resource assignment} and \textit{Time domain resource assignment} specify where the data is located in the resource grid. These parameters do not need to be updated, and the allocation can be done as usual. The decision of whether to use the analog or digital beamforming chain will be implicit, i.e., the assigned resource blocks that fall within the digital beamforming \gls{bwp} will be decoded with digital beamforming. However, there is still the need to add some parameters to the \gls{dci}, as we need to specify a different number of layers and \gls{mcs} for the digital beamforming part. We therefore propose to add the following fields, listed in Tab.~\ref{tab:DCI_1_1_add} to the \gls{dci} \textit{format1\textunderscore1}.

\section{Conclusions}

In this paper, we proposed a novel beamforming architecture, which enables the use of an heterogeneous rank across the bandwidth, i.e. a different number of \gls{rf} chains operating in different portions of the band. We have shown that such an architecture is significantly cheaper and has a much lower power consumption compared with a fully digital beamforming architecture, while maintaining roughly the same performance in a multi-user setting. Moreover, we have shown that such good performance can be achieved with a classical \gls{pf} scheduler, thus making the scheme easy to implement on existing products. Finally, we proposed an update to the \gls{3gpp} standard that would allow for the implementation of such system with few additional parameters.

\bibliographystyle{IEEEtran}  
\bibliography{references}  

%
%
%
%

\vfill

\end{document}